\documentclass[mreferee]{gji}
\usepackage{footnote}
\usepackage{timet}
\usepackage{algorithm}
\usepackage{algorithmic}
\usepackage[pdftex]{graphicx}
\usepackage{epstopdf}
\usepackage{epsfig,subfigure}
\usepackage{epsf}
\usepackage{graphics}
\usepackage{url}
\usepackage{color}
\usepackage{amssymb}
\usepackage{amsmath}

\graphicspath{{./Figures/}}
\addtocounter{footnote}{0}
\newcommand{\bfm}{\mathbf{m}}
\newcommand{\bfd}{\mathbf{d}}

\newcommand{\bfr}{\mathbf{r}}

\newcommand{\bfh}{\mathbf{h}}

\newcommand{\bfu}{\mathbf{u}}
\newcommand{\bfv}{\mathbf{v}}

\newcommand{\tildetildeG}{\tilde{\tilde{G}}}
\newcommand{\Kmax}{K_{\mathrm{max}}}

\newcommand{\WL}{W_{{{L}}_1}}
\newcommand{\Wh}{W_{\mathrm{h}}}
\newcommand{\Wdepth}{W_{\mathrm{depth}}}

\newcommand{\alphaopt}{\alpha_{\mathrm{opt}}}

\newcommand{\sigmamin}{\sigma_{\mathrm{min}}}
\newcommand{\sigmamax}{\sigma_{\mathrm{max}}}
\newcommand{\bfdo}{\mathbf{d}_{\mathrm{obs}}}
\newcommand{\bfdp}{\mathbf{d}_{\mathrm{pre}}}

\newcommand{\bfde}{\mathbf{d}_{\mathrm{exact}}}
\newcommand{\bfma}{\mathbf{m}_{\mathrm{apr}}}

\newcommand{\Wd}{W_{\bfd}}

\newcommand{\Rn}{\mathcal{R}^{n}}
\newcommand{\Rm}{\mathcal{R}^{m}}
\newcommand{\Rmm}{\mathcal{R}^{m \times m}}
\newcommand{\Rnn}{\mathcal{R}^{n \times n}}
\newcommand{\Rmn}{\mathcal{R}^{m \times n}}
\newcommand{\Rmq}{\mathcal{R}^{m \times q}}
\newcommand{\Rqq}{\mathcal{R}^{q \times q}}
\newcommand{\Rnq}{\mathcal{R}^{n \times q}}

\newcommand{\argmin}[1]{\textnormal{arg} \min_{#1}}

\title[Potential field inversions using the RSVD]{Improving the use of the randomized singular value decomposition for the inversion of gravity and magnetic data}
\author[S. Vatankhah, S. Liu, R.~A. Renaut, X. Hu, J. Baniamerian]{Saeed Vatankhah $^1$$^,$$^2$ ,  Shuang Liu $^1$,  Rosemary A. Renaut $^3$, Xiangyun Hu $^1$, \and  Jamaledin Baniamerian  $^4$\\
$^1$ Hubei Subsurface Multi-scale Imaging Key Laboratory, Institute of Geophysics and Geomatics,\\
 China University of Geosciences, Wuhan, China\\
$^2$ Institute of Geophysics, University of Tehran, Tehran, Iran\\
$^3$ School of Mathematical and Statistical Sciences, Arizona State University, Tempe, AZ, USA\\
$^4$ Department of Earth Sciences, College of Sciences and Modern Technologies,\\
Graduate University of Advanced Technology, Kerman, Iran.}
\begin{document}
\maketitle

\begin{summary}
The large-scale focusing inversion of gravity and magnetic potential field data using $L_1$-norm regularization is considered. The use of the randomized singular value decomposition methodology facilitates tackling the computational challenge that arises in the solution of these large-scale inverse problems. As such the powerful randomized singular value decomposition is used for the numerical solution of all linear systems required in the algorithm.  A comprehensive comparison of the developed methodology for the inversion of magnetic and gravity data is presented. These results indicate that  there is generally an important difference between the gravity and magnetic inversion problems. Specifically, the randomized singular value decomposition is dependent on the generation of a rank $q$ approximation to the underlying model matrix, and the results demonstrate that $q$ needs to be larger, for equivalent problem sizes, for the magnetic problem as compared to the gravity problem. Without a relatively large $q$ the dominant singular values of the magnetic model matrix are not well-approximated. The comparison also shows how the use of the power iteration embedded within the randomized algorithm is used to improve the quality of the resulting dominant subspace approximation, especially in magnetic inversion, yielding acceptable approximations for smaller choices of $q$.  The price to pay is the trade-off between approximation accuracy and computational cost. The algorithm is applied for the inversion of   magnetic data obtained  over a portion of the Wuskwatim Lake region in Manitoba, Canada

\end{summary}

\section{Introduction}\label{sec:intro}
Potential field surveys, gravity and magnetic, have been used for many years for a wide range of studies including oil and gas exploration, mining applications, and mapping basement topography \cite{nabighian:2005,Blakely}. The inversion of acquired data is one of the important steps in the interpretation process \cite{LiOl:98,PoZh:99,BoCh:2001,SiBa:2006,Far:2008,Liu:13,Liu:18}. The problem is ill-posed and then, generally, the solution  is obtained via the minimization of a global objective function that consists of  two terms, the data misfit term and the stabilizing, or regularizing, term. These two terms are balanced by a scalar regularization parameter that weights the contribution of the stabilizing term to the solution. Extensive background on the modeling and the solution of the regularized objective function is provided in the literature, e.g. \cite{LiOl:98,PoZh:99,VAR:2015}. While the concept of regularization is well-known, the numerical solution of a large-scale problem, in conjunction with effective and efficient estimation of a suitable regularization parameter, continues to be computationally challenging.  For large-scale problems, the most effective and widely-used strategy is to transform  the problem from the original large space to a much smaller subspace. The resulting subspace solution can then be projected back to the original full space, under reasonable assumptions that the subspace problem sufficiently captures the characteristics of the full space problem. For example the LSQR algorithm, which is based on the Golub-Kahan bidiagonalization of the model matrix, is frequently used for the inversion of geophysical data.  Using the LSQR algorithm the  large-scale problem is projected onto a Krylov subspace of smaller dimension and the subspace  solution is then obtained relatively efficiently using a standard factorization such as the singular value decomposition (SVD), \cite{PaSa:1982a,PaSa:1982b,KiOl:2001,ChNaOl:2008,RVA:2017,VRA:2017}. Still the need to solve ever larger problems so as to provide greater resolution of the subsurface structures, while also automatically estimating a suitable regularization parameter,  presents a challenge computationally. Even with the continually-increasing computational power and memory that is available, it is sometimes impossible, or computationally prohibitive to obtain effective solutions with these traditional computational algorithms.  

Recently, the powerful concept of randomization has been introduced as an alternative strategy for dealing with large-scale inverse problems \cite{Halko:2011,XiZo:2013,VMN:2015,XiZo:2015,WXZ:2016,VRA:2018a,VRA:2018b}. The fundamental idea is that  some amount of randomness can be employed to obtain a matrix of smaller dimension that effectively captures the essential spectral properties  of the original model matrix, and thus provides an approximately optimal  rank-$q$ randomized singular value decomposition (RSVD) of the original system.   Effectively, the approach consists of the first stochastic step which finds a random but orthonormal  matrix that samples rows and columns of a given matrix, and a  second completely deterministic step which then finds the eigen-decomposition of the subsampled matrix and hence yields a rank-$q$ approximation of the original matrix.

For the $3$D inversion of gravity data, Vatankhah et al. \shortcite{VRA:2018b} developed a fast inversion methodology that combined an $L_1$-norm regularization strategy with the RSVD,   for the generation of a  focused image of the subsurface. For the under-determined problem in which there are $m$ data measurements to be used to find the subsurface structures on a volume with $n$  model parameters, $m \ll n$, their results indicate that acceptable results, nearly equivalent to those using the full SVD (FSVD) for the SVD calculations in the algorithm,  are achievable using a target rank  $q$ with {$q \gtrsim (m/6)$}.\footnote[1]{The solutions are the same when $q=m$.} Here we will denote the algorithms that use the RSVD and FSVD, respectively,   for all steps in the $L_1$-norm regularization strategy the \texttt{Hybrid-RSVD} and \texttt{Hybrid-FSVD} algorithms.  For the  \texttt{Hybrid-RSVD} algorithm it is important to estimate an appropriate lower bound on $q$  in order that acceptable solutions are provided using an algorithm that is computationally efficient with respect to both  time  and memory.  

The purpose of this discussion is to assess the application of the \texttt{Hybrid-RSVD} algorithm, as presented in  Vatankhah et al. \shortcite{VRA:2018b}, for the  inversion of magnetic potential data.  Our results demonstrate that direct application of the \texttt{Hybrid-RSVD} algorithm does not lead to acceptable solutions.  Rather, for a problem of equivalent size as given by the pair $(m,n)$,  we find that  $q$  must be larger;  the lower bound $q \gtrsim m/6$ does not yield acceptable solutions. On the other hand, by introducing power iterations into the determination of the rank-$q$ approximation, improved results are achievable for smaller choices of $q$, for both magnetic and gravity inversion problems. Both cases are analyzed carefully and useful estimates for the appropriate choices of $q$ for both magnetic and gravity problems, when solved  using power iterations to improve the approximate RSVD, are provided. 

The remainder of the paper is organized as follows. In Section~\ref{inversionmethod} we briefly describe the standard $L_1$-norm stabilized  inversion algorithm. A short explanation of the methodology used for estimating the regularization parameter is presented in Section~\ref{parameterchoice}. Then, the concept of the RSVD is reviewed in Section~\ref{RSVD}, with the  extension applying power iterations  presented in Section~\ref{powerRSVD}. In Section~\ref{synthetic} we show the results of using the \texttt{Hybrid-RSVD} algorithm on two different synthetic models. A small model in Section~\ref{twodikes} is used for further analysis of the  algorithm and the impact of using  power iterations to obtain the RSVD is also demonstrated. 
To further examine the approach we consider a model with multiple bodies in Section~\ref{multiplebodies}. The inversion of magnetic  data from  a meta-sedimentary gneiss belt in Canada are used to illustrate the application of the presented methodology for real data in Section~\ref{real}. Conclusions and discussions on future work are given in Section~\ref{conclusion}.

\section{ Inversion methodology }\label{inversionmethod}

We discuss the general case for the linear inversion of potential field data in which the subsurface is discretized into a large number of cells of fixed size but with unknown physical properties. The unknown parameters of the cells are stacked in a vector $\bfm \in \Rn$, and  measured potential field data are also  stacked in a vector $\bfdo \in \Rm$. The measurements are connected to the model parameters via the model matrix $G \in \Rmn$, $m\ll n$, which is the forward model operator which maps from model to data spaces, yielding the under-determined linear  system
\begin{eqnarray}\label{d=gm}
\bfdo= G\bfm,
\end{eqnarray}
Element $G_{ij}$   represents the effect of the unit model parameter at  cell $j$ on the data at location $i$. 

In the case of the inversion of magnetic data, vector $\bfm$ collects the values of the unknown susceptibilities of the cells and $\bfdo$ collects the total magnetic field.  For the gravity problem these are the cell densities and vertical component of gravity field, respectively. Model matrix $G$ depends on the problem.  Rao  $\&$ Babu \shortcite{RaBa:91} provided a  fast approach for computing the total magnetic field anomaly of a cube that is then  used here to form the elements of model matrix $G$ for the magnetic problem. For gravity inversion,  the elements of $G$ are computed using the formula developed by Ha{\'a}z \shortcite{Haaz}, see for example Boulanger \& Chouteau \shortcite{BoCh:2001} for more details. The spectral properties of these two matrices impact the condition of the under-determined systems, and hence the performance of any algorithm that is used for data inversion.

Stabilization, or regularization, is required to find an acceptable solution of  the ill-posed system given by  \eqref{d=gm}; its solution is neither unique nor stable. Here we consider the $L_1$-norm stabilized  solution of \eqref{d=gm} as  presented by Vatankhah et al. \shortcite{VRA:2017}. The approach also includes weighting of the data based on the knowledge, or estimate, of the standard deviation of the independent noise in the data,  weighting for the depths of the cells, and the inclusion of prior  knowledge on the solution using provided information  which may come from geology, logging or previous geophysical surveys, or may be taken to be the vector of $0$ values when no prior information is available. Overall, the formulation finds the minimum of the global objective function $P^{\alpha}(\bfm)$ as given by
\begin{eqnarray}\label{globalfunction1}
\bfm = \argmin{\bfm} \{P^{\alpha}(\bfm)\} =\argmin{\bfm}\{\| \Wd(G\bfm-\bfdo)  \|_2^2 + \alpha^2 \|W(\bfm-\bfma) \|_2^2\}.
\end{eqnarray} 
The diagonal entries of the data weighting matrix $\Wd$ are estimates for the  inverse of the standard deviations of the  independent noise in the data, $\bfma$ encodes the prior knowledge of the solution, and stabilizing matrix  $W$ is the product of three diagonal matrices, $\Wdepth$ which is a depth-weighting matrix with diagonal entries $z^{-\beta}$ at depth $z$, $\Wh$ which is a hard constraint matrix, and $\WL$ which is a matrix that arises from the approximation of the $L_1$-norm stabilizer via an $L_2$-norm term. Details of all aforementioned matrices are provided in Vatankhah et al. \shortcite{VRA:2017,VRA:2018b}, but we note that it is $\WL$ which enables the algorithm to produce non-smooth and focused images of the subsurface that are more consistent with real geological structures.  
The scalar regularization parameter $\alpha$ balances the two terms in the objective function and is discussed further in Section~\ref{parameterchoice}

Noting that the inverse of a diagonal matrix is obtained at effectively zero computational cost, the objective function \eqref{globalfunction1}
is easily transformed to the standard Tikhonov form 
\begin{eqnarray}\label{globalfunction2}
P^{\alpha}(\bfh)=\| \tildetildeG\bfh - \tilde{\bfr}  \|_2^2 + \alpha^2 \| \bfh \|_2^2,
\end{eqnarray}
see e.g. Vatankhah et al. \shortcite{VAR:2015}, where $\tildetildeG={\Wd}GW^{-1}$, $\tilde{\bfr}={\Wd}(\bfdo-G\bfma)$, and $\bfh=W(\bfm-\bfma)$. The solution of \eqref{globalfunction2}, dependent on the choice of $\alpha$, is then
\begin{eqnarray}\label{hsolution}
\bfh({\alpha})=(\tildetildeG^T\tildetildeG+\alpha^2 I_n)^{-1}\tildetildeG^T\tilde{\bfr}, 
\end{eqnarray}
and the model update is given by
\begin{eqnarray}\label{modelupdate}
\bfm(\alpha)=\bfma+W^{-1}\bfh (\alpha).
\end{eqnarray}
An iteratively reweighted approach is used to find  $\bfm(\alpha)$, dependent not only on $\alpha$, changing with each iteration, but also on matrix $W$ which changes at each step through the update of matrix $\WL$.  The complete approach is described in  Vatankhah et al. \shortcite[Algorithm~$1$]{VRA:2017} and Vatankhah et al. \shortcite[Algorithm~$2$]{VRA:2018b}. At each stage of the algorithm upper and lower bounds on the physical parameters are imposed in order that the recovered model is reliable within known acceptable ranges, and regularization parameter $\alpha$ is adjusted automatically. The iteration is terminated when either the  data predicted by the reconstructed model satisfies the observed data to within a $\chi^2$ value relating to the noise level, or a maximum number of iterations $\Kmax$  is reached without the predicated data satisfying  the $\chi^2$ estimate. 

When both $m$ and $n$ are small,  $\bfh(\alpha)$ can be found  using the FSVD $\tildetildeG = U \Sigma V^T$, where matrices $U\in \Rmm$ and $V\in\Rnn$ are orthogonal with columns $\bfu_i$ and $\bfv_i$, and $\Sigma \in \Rmn$ is the matrix of singular values $\sigma_i$, ordered from large to small. The solution obtained using the  FSVD is  described in Vatankhah et al \shortcite[Algorithm~$1$]{VRA:2017}, and see also Paoletti et al. \shortcite{PHHF:2014} and  Vatankhah et al. \shortcite{VAR:2015}. The availability of the FSVD makes it  possible to estimate regularization parameter $\alpha$ cheaply using standard parameter-choice techniques \cite{XiZo:2013,ChPa:2015}. Unfortunately the calculation of the  SVD for large, or even moderate, under-determined systems is not practical;  the cost is approximately $6nm^{2}+20 m^{3}$, Golub $\&$ Van Loan \shortcite{GoLo:2013}. The traditional alternative is the use of a hybrid method such as the iterative LSQR algorithm that can be used to project \eqref{globalfunction2} onto a Krylov subspace of smaller dimension and for which an SVD of the projected problem is then efficiently used to yield the subspace solution. This solution is projected back to the original full space at minimal additional computational cost. The SVD for the projected problem also facilitates the use of parameter-choice algorithms to find an optimal $\alpha$, denoted by $\alphaopt$,  also at minimal additional computational cost. 

\subsubsection{Regularization parameter-choice method}\label{parameterchoice}
Here  we use the method of unbiased predictive risk estimation (UPRE) to find $\alphaopt$. This a-posteriori rule for choosing the Tikhonov regularization parameter method is well-described in Vogel \shortcite{Vogel:2002}, and has been extensively applied for the inversion of data when  an estimate of the noise covariance is available, including for the inversion of geophysical data, \cite{VAR:2015,VRA:2017} and for more general inversion problems \cite{RVA:2017,ChPa:2015}. Using the SVD of matrix $\tildetildeG$, the UPRE function to be minimized is given by
\begin{eqnarray}\label{upresvd}
U(\alpha)=\sum_{i=1}^{m^*}  \left( \frac{1}{\sigma_i^2 \alpha^{-2} + 1} \right)^2 \left(\bfu_i^T\tilde{\bfr} \right)^2 + 2 \left( \sum_{i=1}^{m^*} \frac{\sigma_i^2}{\sigma_i^2+\alpha^2}\right) - {m^*}.
\end{eqnarray}  
Here, ${m^*}$ indicates the number of non-zero singular values. This is the numerical rank of the matrix, namely $m^*=m$ when UPRE is applied using the FSVD for the underdetermined problem. The optimum regularization parameter, $\alphaopt$, is found by evaluating \eqref{upresvd} on a range of $\alpha$, between minimum and maximum $\sigma_i$; equivalently $\alphaopt=\argmin{\sigmamin\le  \alpha \le\sigmamax}\{U(\alpha)\}$.

\subsection{Randomized Singular Value Decomposition}\label{RSVD}
While the hybrid-LSQR methodology is effective and practical; acceptable results are obtained as compared with the \texttt{Hybrid-FSVD} solution, see for example Renaut et al. \shortcite{RVA:2017} and Vatankhah et al. \shortcite{VRA:2017},  the approach is still cost-limited due to the need to build a relatively large Krylov space for the solution when both $m$ and $n$ are large. Recent approaches based on randomization have presented an interesting alternative for tackling problems requiring high resolution of the subsurface,  \cite{XiZo:2013,VMN:2015,VRA:2018a,VRA:2018b}. In this case, random sampling is used to construct a low-dimensional subspace that approximates the column space of the model matrix and maintains  the most dominant spectrum of the original matrix, \cite{Halko:2011}. Standard deterministic matrix decomposition methods such as the SVD, or eigen-decomposition, can then be used to compute a low-rank approximation of the original matrix. Specifically, in the context of potential field inversion, it is desirable to find a $q$-rank matrix $\tildetildeG_q$, which is as close as possible to  $\tildetildeG$ in the least-squares sense, while at the same time the target rank $q$ is as small as possible in order that the inversion process is fast. We note that of course the best rank $q$ approximation in the least squares sense is given by the exact truncated SVD of $\tildetildeG$ with $q$ terms \cite{GoLo:2013}. It is generally, however, not practical to calculate the truncated SVD for  large scale problems. Thus here we focus on the RSVD and carefully describe the approach that was presented in  Vatankhah et al \shortcite{VRA:2018b} for the solution of the gravity inversion problem. 

The fundamental aspects of an RSVD algorithm consist of two stages. Here we present this for the under-determined matrix $\tildetildeG$.  (i) A low-dimensional subspace is constructed that approximates the column space of $\tildetildeG^T$. The aim is  to find a matrix $Q \in \mathcal{R}^{n \times q}$ with orthonormal columns such that $\tildetildeG \approx \tildetildeG Q Q^T$; (ii) Given the near-optimal basis spanned by the columns of $Q$, a smaller matrix $B=\tildetildeG Q \in \mathcal{R}^{m \times q}$ is formed. This means that $\tildetildeG$ is restricted to the smaller subspace spanned by  the basis from the columns of $Q$. Moreover, $B$  is a projection from high-dimensional space into a low-dimensional space which preserves the geometric structure of the matrix in a Euclidean sense \cite{Eri:2016}. Subsequently, $B$ can then be used to compute an approximate  matrix decomposition for $\tildetildeG$ using a traditional algorithm. Step (i) is completely random and depends on the selection of a specific approach to find $Q$, while (ii) is deterministic. The fundamental approach, as presented for the gravity problem by  Vatankhah et al.  \shortcite[Algorithm~$1$]{VRA:2018b}, as summarized in Algorithm~\ref{RSVDAlgorithm}, but with the inclusion of power iterations,  is now discussed.

In step~\ref{omega} a random test matrix $\Omega \in \mathcal{R}^{\ell \times m} $ is generated from a standard normal distribution. Probability theory guarantees that the columns of $\Omega$ are linearly independent. Here, $q+p=\ell \ll m$ and $p$ is a small oversampling parameter  that provides the flexibility and effectiveness of the algorithm \cite{Halko:2011,Eri:2016}.  At step~\ref{Y} a set of independent randomly-weighted linear combinations of the rows of $\tildetildeG$, or equivalently columns of $\tildetildeG^T$, are formed. The sketch matrix $Y$ has a much smaller number of rows than $\tildetildeG$. Ignoring for the moment power iterations at step~\ref{powerstep},  step~\ref{Q} constructs  $Q \in \mathcal{R}^{n \times \ell}$. The columns of $Q$ form an orthonormal basis for the range of $Y^T$; equivalently for the  range of $\tildetildeG^T$. Given near-optimal $Q$, a smaller matrix $B$ is reconstructed in step~\ref{B}. Therefore, large matrix $\tildetildeG$ is projected onto a low-dimensional space that captures most of the action of $\tildetildeG$. Choosing an optimal $q$ is highly dependent on the task. Generally, $q$ should be as small as possible so that the algorithm is fast and efficient, but simultaneously $q$ should be large enough that the dominant spectral properties of $\tildetildeG$ are accurately captured. We discuss the effect of the choice of $q$ for the inversion of  gravity and magnetic data in Section~\ref{synthetic}.

Having obtained matrix $B$ at step~\ref{B}, traditional SVD algorithms could then be used  to compute the approximations to the first $q$ left singular vectors as well as the corresponding singular values for matrix $\tildetildeG$. The approximate right singular vectors could also be recovered,  see e.g. Xiang and Zou \shortcite{XiZo:2013}. Alternatively, as discussed, with proof, in Vatankhah et al. \shortcite{VRA:2018b}, the much smaller matrix $B^TB\in \mathcal{R}^{\ell \times \ell}$ can be used to find the required SVD components for $B$ using the eigen-decomposition of $ B^TB$. It was also demonstrated that the computational cost of this algorithm, excluding power iterations,  is  $O(\ell mn)$. Thus, it is feasible to compute the large singular values of  a given matrix efficiently.

\begin{algorithm}
\caption{RSVD algorithm with power iterations. Given matrix $\tildetildeG \in \Rmn (m < n)$, a target matrix rank $q $ and a small constant oversampling parameter $p$ satisfying $q+p=\ell \ll m$, compute an approximate SVD of $\tildetildeG$: $\tildetildeG$ $\approx$  $ U_q \Sigma_q V_q^T$ with $U_q \in \Rmq$,  $\Sigma_q \in \Rqq$, $V_q \in \Rnq$.}\label{RSVDAlgorithm}
\begin{algorithmic}[1]
\STATE Generate a Gaussian random matrix $\Omega \in \mathcal{R}^{\ell \times m} $. \label{omega}
\STATE Form the matrix $Y=\Omega \tildetildeG \in \mathcal{R}^{\ell \times n}$. \label{Y}
\STATE Compute power scheme, see Algorithm~\ref{SubPowerAlgorithm}.\label{powerstep} 
\STATE Compute orthonormal matrix $Q \in \mathcal{R}^{n \times \ell}$ via QR factorization $Y^T=QR$.\label{Q}
\STATE Form the matrix $ B=\tildetildeG Q \in \mathcal{R}^{m \times \ell}$.\label{B}
\STATE Compute the matrix $B^TB \in \mathcal{R}^{\ell \times \ell}$.\label{BTB}
\STATE Compute the eigendecomposition of $B^TB$; $[\tilde{V}_\ell, D_\ell]=\mathrm{eig}(B^TB)$. \label{eigen}
\STATE Compute $V_q=Q \tilde{V}_\ell(:,1:q)$;  $\Sigma_q= \sqrt{D_\ell}(1:q,1:q)$; $\mathrm{and}$ $ U_q=B \tilde{V}_q(:,1:q)$ $ \Sigma_{q}^{-1}. $\label{SVDcomponent}
\STATE Note $\tildetildeG_q = U_q \Sigma_q V_q^T$ is a q-rank approximation of matrix $\tildetildeG$.
\end{algorithmic}
\end{algorithm}

\subsubsection{Randomized Singular Value Decomposition with Power Iterations}\label{powerRSVD}
The quality of the RSVD which is obtained using Algorithm~\ref{RSVDAlgorithm} without step~\ref{powerstep} depends on the quality of the basis matrix $Q$ as providing a basis for the column space of $\tildetildeG^{T}$. Halko et al \shortcite[p224]{Halko:2011}  suggested an improvement of their \texttt{Proto} Algorithm for forming the basis matrix $Q$ and the associated RSVD that should lead to an improved approximation of the dominant spectrum. In the \texttt{Prototype} Algorithm for power iterations \cite[p227]{Halko:2011}, the sketch matrix $Y$ is obtained after  first preprocessing   matrix $\tildetildeG$ to give 
\begin{eqnarray}\label{power}
\tildetildeG^{(s)}=\tildetildeG(\tildetildeG^T\tildetildeG)^{(s)},
\end{eqnarray}
where integer $s$ specifies the number of power iterations. While this is specifically implemented as \cite[Algorithm 4.3]{Halko:2011}, Halko et al \shortcite{Halko:2011} note that the approach is sensitive to floating point rounding errors, which reduces the quality of the $Q$ basis. Thus, Halko et al \shortcite{Halko:2011} suggested their \texttt{Algorithm 4.4} which orthonormalizes the columns of $Y$ between each application of $\tildetildeG$ and $\tildetildeG^T$. It is this latter approach, described in Algorithm~\ref{SubPowerAlgorithm},  that we implement for step~\ref{powerstep} within Algorithm~\ref{RSVDAlgorithm},  but here with our extension of the approach for the under-determined case.

\begin{algorithm}
\caption{Subspace iteration of power scheme. For input matrix $\tildetildeG \in \Rmn (m < n)$, the sketch $Y \in \mathcal{R}^{\ell \times n}$, and parameter $s$, return an improved sketch matrix.}\label{SubPowerAlgorithm}
\begin{algorithmic}[1]
\STATE For $j=1,\cdots, s$.
\STATE $[Q,\sim]=qr(Y^{T},0)$. (economic QR decomposition)
\STATE $[Q,\sim]=qr(\tildetildeG Q,0)$.
\STATE $Y^T=\tildetildeG^T Q$.
\STATE End 
\end{algorithmic}
\end{algorithm}

To analyze the effect of power iterations on improving the accuracy of the computed matrix, a simplified upper bound on the expected error between the original and the $q$-rank computed matrices is given by
\begin{eqnarray}\label{expectederror}
E \| \tildetildeG- \tildetildeG_q \| \leq \left[1+\sqrt{\frac{q}{p-1}}+\frac{e\sqrt{q+p}}{p}.\sqrt{min(m,n)-q}  \right]^{\frac{1}{2s+1}} \sigma_{q+1}, 
\end{eqnarray}
 \cite{Martinsson:2016,Eri:2016}. Here $e$ is Euler's number, $\sigma_{q+1}$ is the $q+1$ largest singular value of the matrix 
$\tildetildeG$, $E$ denotes the expectation operator, and it is assumed that $p\ge 2$. Upper bound \eqref{expectederror} indicates  how parameters $p$, $q$, and $s$ can be used to control the approximation error.  Note immediately that for $q=min(m,n)$  then $\sigma_{q+1}=0$ and  $E \| \tildetildeG- \tildetildeG_q \|=0$.  With increasing oversampling $p$, the second and third terms in the bracket tend toward zero which means that the bound approaches  the theoretically optimum value $\sigma_{q+1}$ \cite{Eri:2016}. For larger values of the subspace iteration parameter  $s$, $1/(2s+1)$ goes to zero and      the error bound is reduced. 

 In Section~\ref{synthetic} we will illustrate the application of   Algorithm~\ref{RSVDAlgorithm} for the inversion of synthetic  gravity and magnetic data sets without power iterations at step~\ref{powerstep}. These results, especially for the magnetic data, suggest that power iterations as indicated in Algorithm~\ref{SubPowerAlgorithm} are needed to improve the approximation of the dominant spectral space.   Our tests have shown that setting $s=1$ yields a good performance that trades-off between accuracy and computational time. We will see that the magnitude of the singular values of the magnetic kernel are larger for a given $q$ than their counterparts for the gravity kernel, and thus the upper bound estimate in \eqref{expectederror} for the approximation error is consistently higher for the magnetic problem. Equivalently this leads to the need to implement a power iteration scheme to reduce the factor multiplying $\sigma_{q+1}$ in \eqref{expectederror}.

\section{Synthetic examples}\label{synthetic}
In Section~\ref{twodikes}  we first evaluate the performance of the RSVD algorithm without power iterations, comparing its performance for the solution of  relatively small-scale gravity and magnetic problems under the same configuration, but the appropriate choice of model matrix $\tildetildeG$. The results are compared to those obtained using the \texttt{Hybrid-FSVD} in each case. To understand the performance of the algorithm we examine the spectrum of the approximate operator, as compared to that of $\tildetildeG$, for each problem in Section~\ref{sec:spectrum}, and then examine the improvement obtained using  the power iterations for $s=1$ in Section~\ref{poweriteration}. A more complicated configuration for a structure with multiple bodies is then examined in  Section~\ref{multiplebodies} and confirms the conclusions obtained for the dipping dike example in Section~\ref{twodikes}.

In the simulations  for the generation of the total field anomaly,  the intensity of the geomagnetic field, the inclination, and the declination are selected as $47000$ nT, $50^{\circ}$ and $2^{\circ}$, respectively. The density contrast  and the susceptibilities of the model structures embedded in a homogeneous non-susceptible background are  $\rho=1$~g~cm$^{-3}$ and $\kappa=0.1$ (SI unit), respectively.  The bound constraints  $0=\rho_{\mathrm{min}}\le \rho \le \rho_{\mathrm{max}}=1$ in units ~g~cm$^{-3}$ and $0=\kappa_{\mathrm{min}}\le \kappa \le \kappa_{\mathrm{max}}=0.1$ in SI units are imposed at each iteration of the gravity and magnetic inversions, respectively. Further,  for all simulations  we   add Gaussian noise with zero mean and standard deviation $ (\tau_1~|\bfde|_i + \tau_2~ \mathrm{max}| \bfde |)$ to datum $i$, for chosen pairs $(\tau_1, \tau_2)$, where $\bfde$ is the exact data set, yielding $\bfdo$ with a known distribution of noise for the error. This standard deviation is used to generate the  matrix $\Wd$ in the data fit term. The values of 
$(\tau_1,\tau_2)$ are selected such that the signal to noise ratios, as given by 
\begin{align}\label{snr}
SNR=20\, \mathrm{log}_{10} \frac{\| \bfde \|_2}{\| \bfdo-\bfde \|_2},
\end{align}
are close for both gravity and magnetic noise-contaminated data. The values for $(\tau_1, \tau_2)$ and the resulting SNRs are specified in the captions of figures associated with the results for each data set.  Then to test convergence of the update $\bfm^{(k)}$ at iteration $k$ we calculate the $\chi^2$ estimate, 
\begin{align}\label{chi2}
(\chi^2)^{(k)}=\|  \Wd(\bfdo-\bfdp^{(k)})\|_2^2, 
\end{align} 
where $\bfdp^{(k)} = G\bfm^{(k)}$, and which assesses the predictive capability of the current solution. When $(\chi^2)^{(k)} \leq m+\sqrt{2m}$ the iteration terminates. Otherwise, the iteration is allowed to proceed in all cases to a maximum number of iterations $\Kmax=50$.   The regularization parameter $\alpha$ is adjusted with iteration $k$ and for $k>1$ is obtained in all cases using the UPRE function \eqref{upresvd} with the appropriate approximate SVD terms. Based on the suggestion of Farquharson \& Oldenburg \shortcite{Far:2004}, a large $\alpha$ is always used at the first iteration, 
\begin{align*}
\alpha^{(1)} = \left(\frac{n}{m}\right)^{3.5} \frac{\sigma_1}{\mathrm{mean}(\sigma_i)},
\end{align*} 
as used in Vatankhah et al. \shortcite[eq.19]{VRA:2017}. In all simulations we use a fixed oversampling parameter, $p=10$, assume $\bfma=\mathbf{0}$, impose physically reasonable constraints on the model parameters, and for $\Wdepth$  take $\beta=0.8$ and $1.4$,  for gravity and magnetic inversions, respectively,  with values close to those suggested by Li $\&$ Oldenburg \shortcite{LiOl:98}. In the results we examine the dependence of the solution on the choice of $q$ for the rank $q$ approximation and record (i) the number of iterations $K$ required, (ii) the relative error  progression with increasing $k$ as given by 
\begin{align}\label{RE}
RE^{(k)}=\frac{\|\bfm_{\mathrm{exact}}-\bfm ^{(k)} \|_2}{\|\bfm_{\mathrm{exact}} \|_2}
\end{align}
(iii) the relative error in the rank $q$ approximation to $\tildetildeG$, 
\begin{align}\label{RE}
RG^{(k)}=\frac{\|\tildetildeG-\tildetildeG_q^{(k)} \|_2}{\|\tildetildeG \|_2}
\end{align}
(iv) $\alpha^{(k)}$ with increasing $k$, and (v) all values at the final iteration $K$ as well as the time for the iterations. 
 Moreover, in recording the computational time to convergence, in all cases we do not include the calculation of the original model matrix $G$.
Computations are performed on a desktop computer with Intel(R) Xeon(R) W-2133 CPU 3.6~GHz processor and 32 GB RAM.

 \subsection{Small-scale model consisting of two dipping dikes}\label{twodikes}
\begin{figure*} 
\subfigure{\label{fig1a}\includegraphics[width=.45\textwidth]{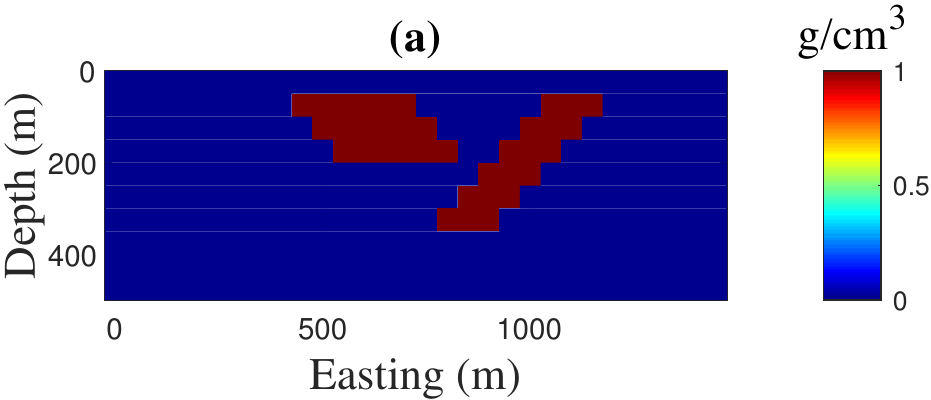}}
\subfigure{\label{fig1b}\includegraphics[width=.45\textwidth]{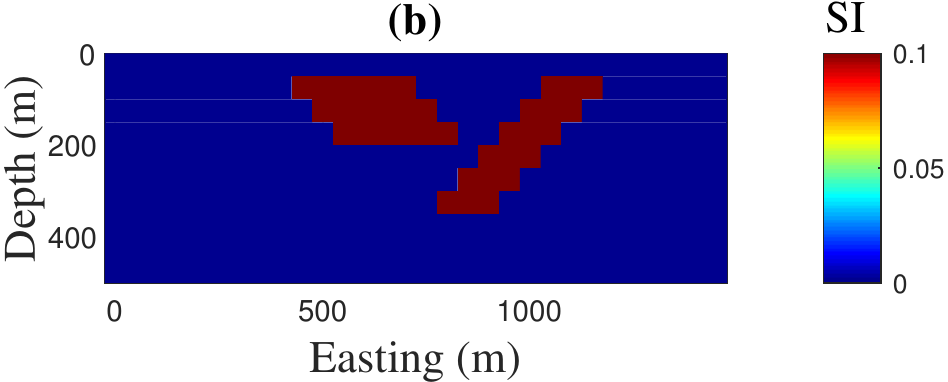}}
\caption {Cross-section of the synthetic model consisting of two dipping dikes. (a) Density distribution; (b) Susceptibility distribution.} \label{fig1}
\end{figure*}

\begin{figure*} 
\subfigure{\label{fig2a}\includegraphics[width=.40\textwidth]{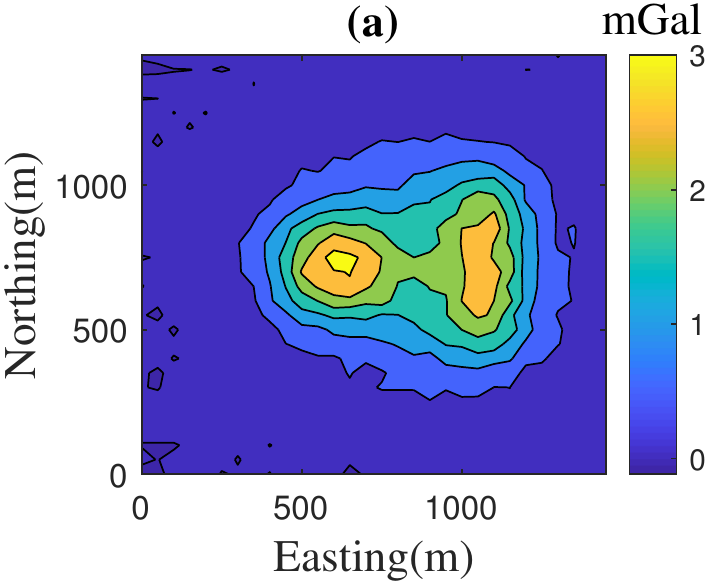}}
\subfigure{\label{fig2b}\includegraphics[width=.43\textwidth]{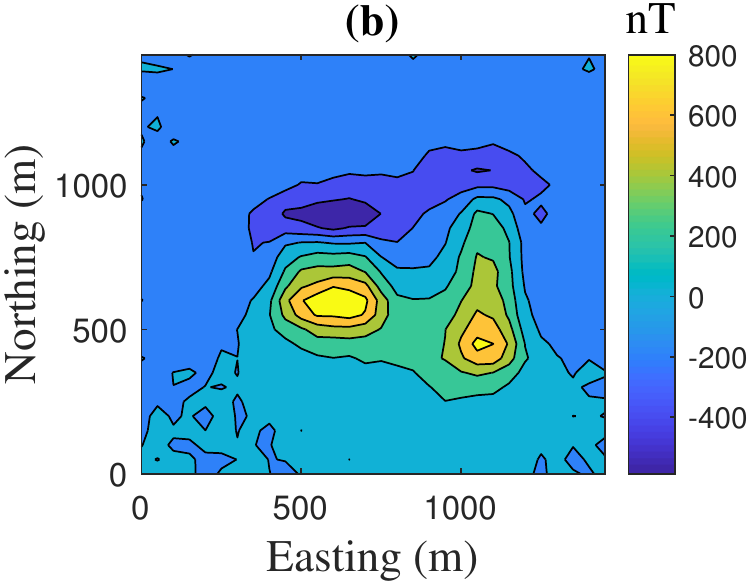}}
\caption {Anomaly produced by the model shown in Fig.~\ref{fig1} and contaminated by Gaussian noise. (a) Vertical component of the gravity field. The noise is generated using parameter pairs $(\tau_1=0.02, \tau_2=0.02)$; (b) Total magnetic field. Here, $(\tau_1=0.02, \tau_2=0.015)$. The $\mathrm{SNR}$ for  gravity and magnetic data, respectively, are $21.9188$ and $21.7765$.} \label{fig2}
\end{figure*}

The small-scale but complicated structure of two dipping dikes, illustrated in Fig.~\ref{fig1}, makes it computationally feasible to compare the solutions for the gravity and magnetic inverse problems using Algorithm~\ref{RSVDAlgorithm} without power iteration, with the solutions obtained using the \texttt{Hybrid-FSVD}. 
 The data for the problem, the vertical component of the gravity and the total magnetic field, are generated on the surface on   $30 \times 30 = 900$ grid  with grid spacing $50$~m.  The noisy gravity and magnetic data in each case are illustrated in Figs.~\ref{fig2a} and \ref{fig2b}. 

\begin{figure*} 
\subfigure{\label{fig3a}\includegraphics[width=.45\textwidth]{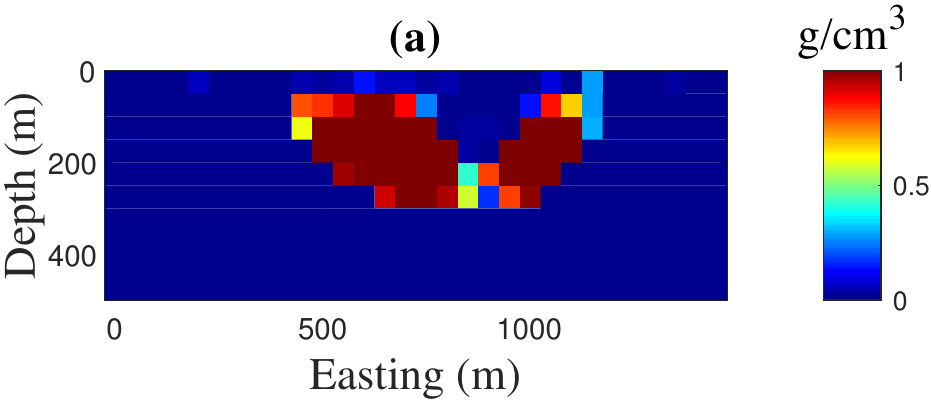}}
\subfigure{\label{fig3b}\includegraphics[width=.45\textwidth]{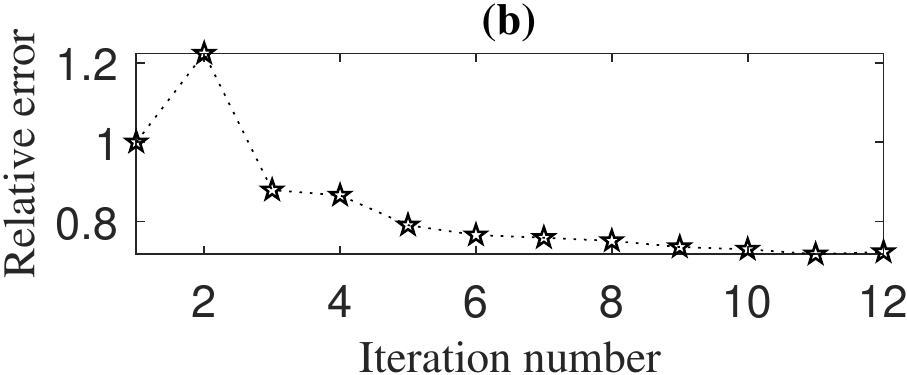}}
\subfigure{\label{fig3c}\includegraphics[width=.45\textwidth]{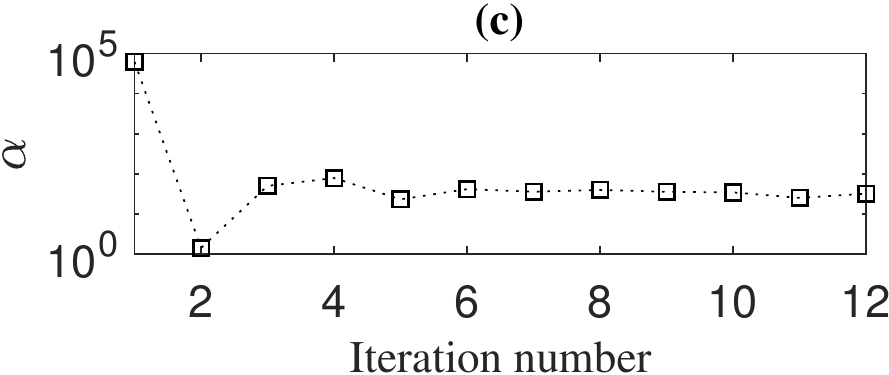}}
\subfigure{\label{fig3d}\includegraphics[width=.45\textwidth]{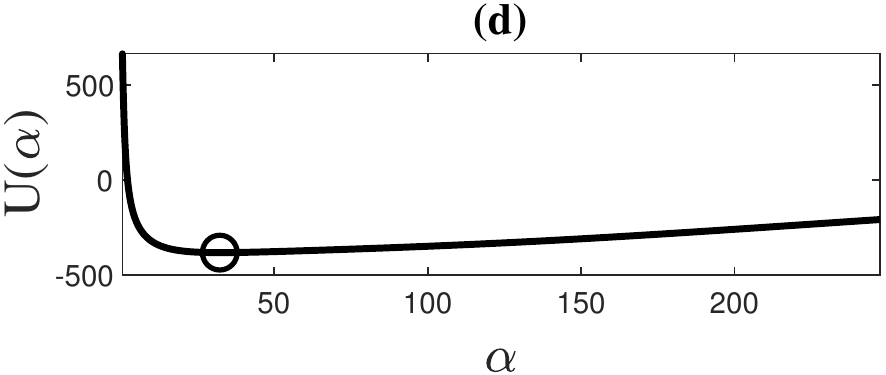}}
\caption {\texttt{Hybrid-FSVD} results for the inversion of gravity data given in Fig.~\ref{fig2a}. (a) Cross-section of reconstructed model; (b) The progression of relative error $RE^{(k)}$ with iteration $k$; (c) The progression of regularization parameter $\alpha^{(k)}$ with iteration $k$; (d) The UPRE function at the final iteration.} \label{fig3}
\end{figure*}

\begin{figure*} 
\subfigure{\label{fig4a}\includegraphics[width=.45\textwidth]{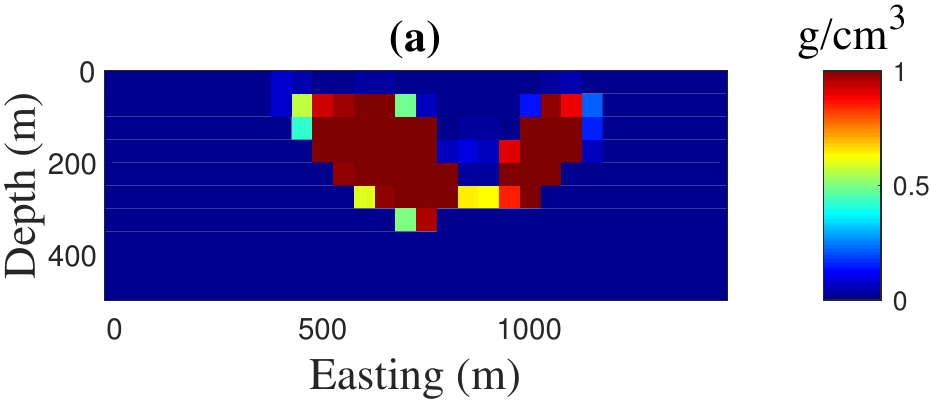}}
\subfigure{\label{fig4b}\includegraphics[width=.45\textwidth]{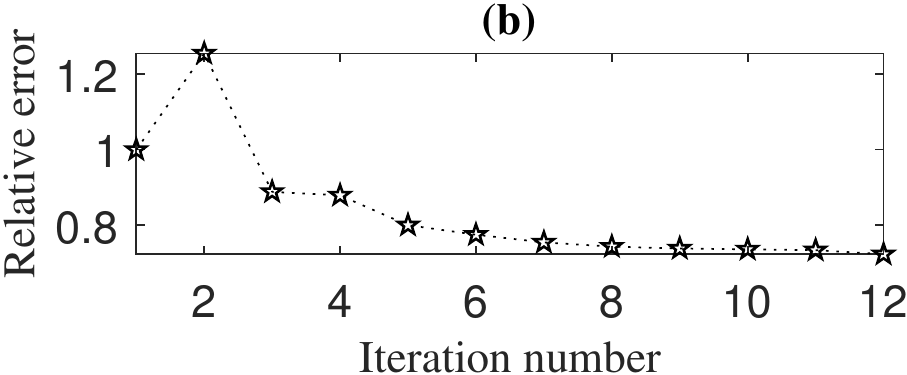}}
\subfigure{\label{fig4c}\includegraphics[width=.45\textwidth]{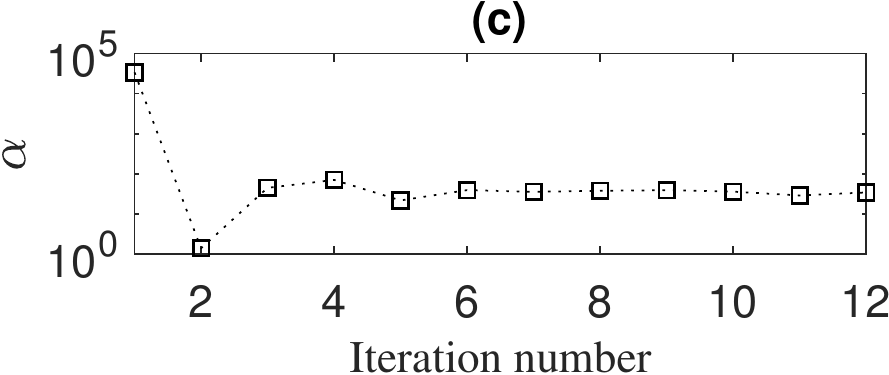}}
\subfigure{\label{fig4d}\includegraphics[width=.45\textwidth]{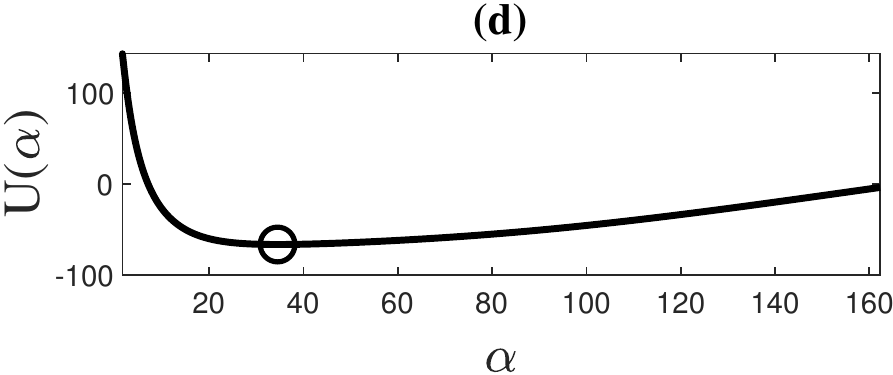}}
\caption {\texttt{Hybrid-RSVD} results using target rank $q=200$ for the inversion of gravity data given in Fig.~\ref{fig2a}.  (a) Cross-section of reconstructed model; (b) The progression of relative error $RE^{(k)}$ with iteration $k$; (c) The progression of regularization parameter $\alpha^{(k)}$ with iteration $k$; (d) The UPRE function at the final iteration.} \label{fig4}
\end{figure*}

\begin{figure*} 
\subfigure{\label{fig5a}\includegraphics[width=.45\textwidth]{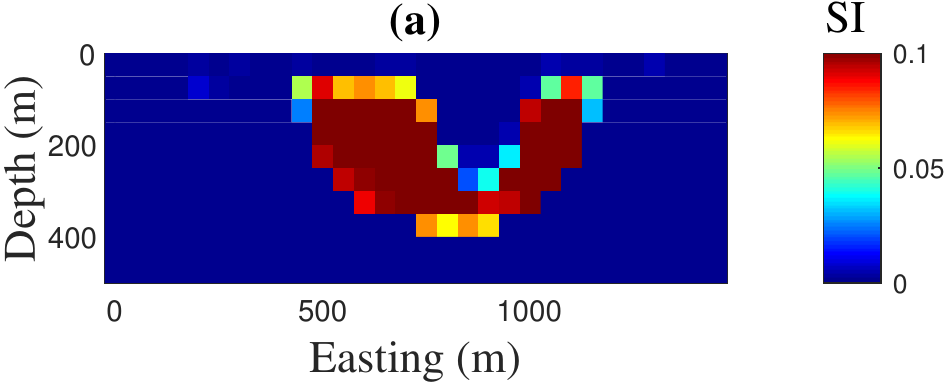}}
\subfigure{\label{fig5b}\includegraphics[width=.45\textwidth]{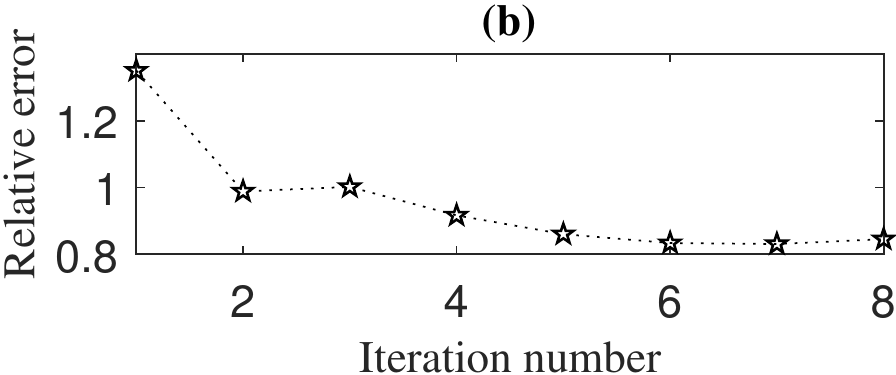}}
\subfigure{\label{fig5c}\includegraphics[width=.45\textwidth]{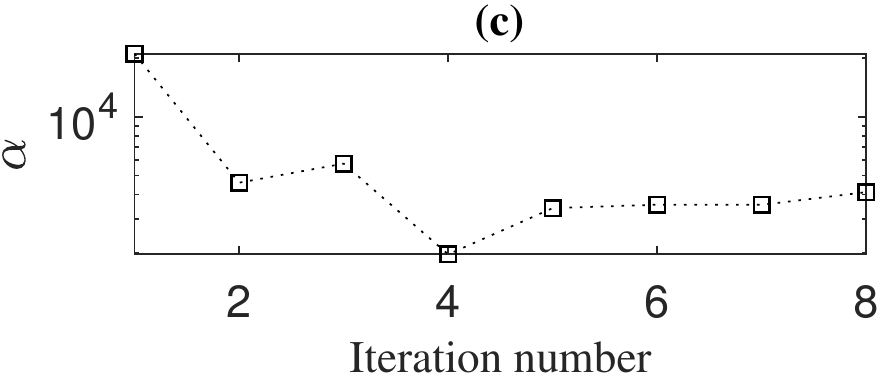}}
\subfigure{\label{fig5d}\includegraphics[width=.45\textwidth]{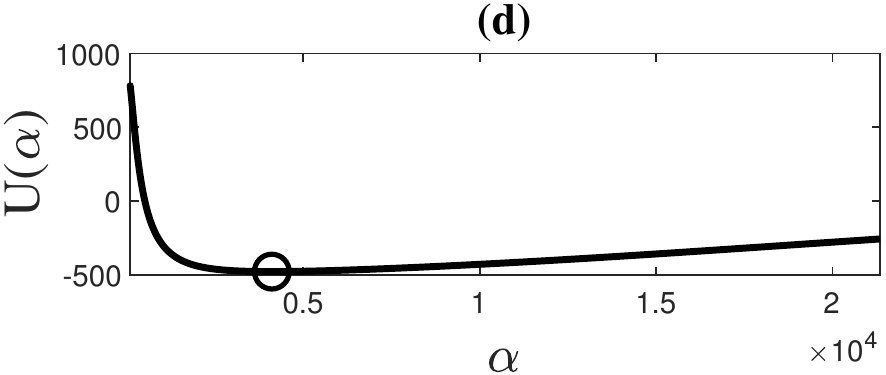}}
\caption {\texttt{Hybrid-FSVD} results for the  inversion of magnetic data given in Fig.~\ref{fig2b}. (a) Cross-section of reconstructed model; (b) The progression of relative error $RE^{(k)}$ with iteration $k$; (c) The progression of regularization parameter $\alpha^{(k)}$ with iteration $k$; (d) The UPRE function at the final iteration.} \label{fig5}
\end{figure*}

\begin{figure*} 
\subfigure{\label{fig6a}\includegraphics[width=.45\textwidth]{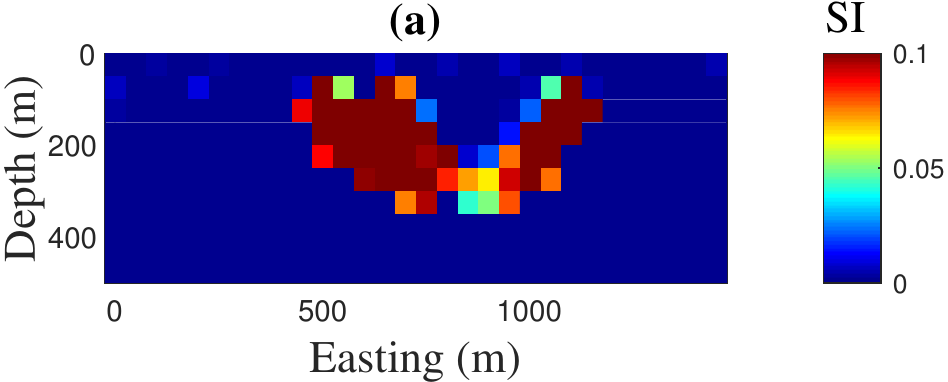}}
\subfigure{\label{fig6b}\includegraphics[width=.45\textwidth]{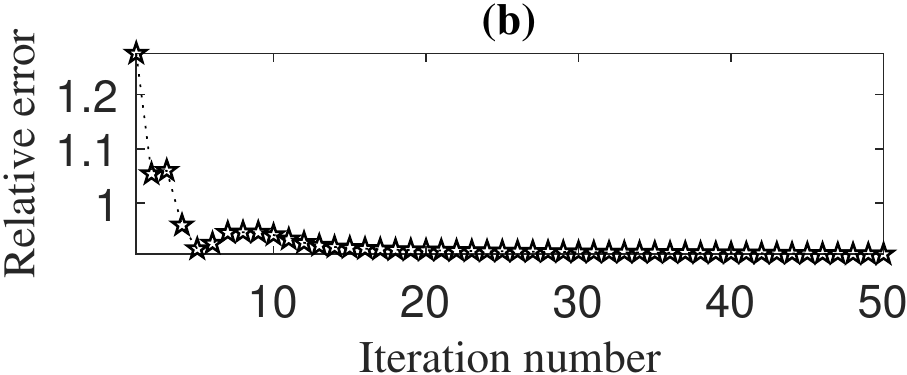}}
\subfigure{\label{fig6c}\includegraphics[width=.45\textwidth]{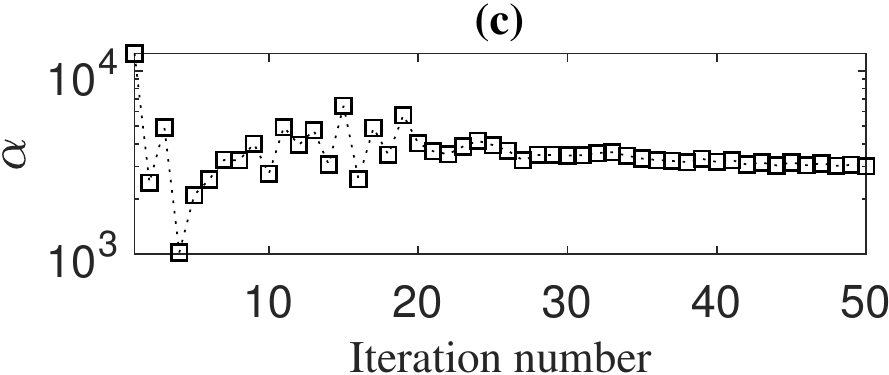}}
\subfigure{\label{fig6d}\includegraphics[width=.45\textwidth]{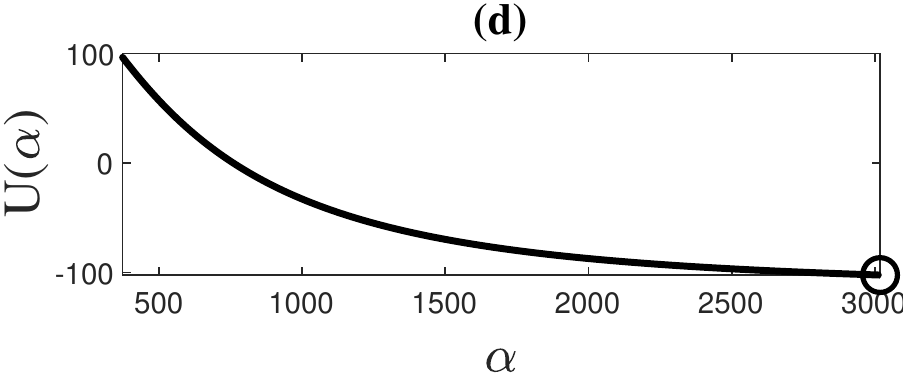}}
\caption {\texttt{Hybrid-RSVD} results using target rank $q=200$ for the inversion of magnetic data given in Fig.~\ref{fig2b}.  (a) Cross-section of reconstructed model; (b) The progression of relative error $RE^{(k)}$ with iteration $k$; (c) The progression of regularization parameter $\alpha^{(k)}$ with iteration $k$; (d) The UPRE function at the final iteration.} \label{fig6}
\end{figure*}

The subsurface volume is discretized into $9000$ cubes of size $50$~m in each dimension, corresponding to $10$ slices in depth for a cross section of $30$ by $30$ cells. The resulting matrix $\tildetildeG$  is of size $900 \times 9000$ and thus facilitates solutions using the  FSVD  for comparison with the RSVD in the inversion algorithm.  The results obtained using the \texttt{Hybrid-FSVD} and \texttt{Hybrid-RSVD}, with the choices $q=100$, $150$, $200$, $300$, $500$, $700$, and $900$, are detailed in Table~\ref{gravitytab} and Table~\ref{magnetictab}, respectively. The results for the  inversions using the \texttt{Hybrid-FSVD} for the gravity and magnetic data inversions are given in Fig.~\ref{fig3} and Fig.~\ref{fig5}, respectively. For comparison the results using $q=200$ are illustrated in Fig.~\ref{fig4} and Fig.~\ref{fig6} for the gravity inversion and  magnetic data inversion, respectively.

The results presented for the inversion of the gravity data using the \texttt{Hybrid-FSVD} show that the iteration terminates after just $12$ iterations, and as indicated in Fig.~\ref{fig3a}  the reconstructed model is in good agreement with the original model. A sharp and focused image of the subsurface is obtained, and while the depths to the top of the structures are consistent with those of the original model, the extensions of the dikes  are overestimated for the left dike and underestimated for the right dike. Figs.~\ref{fig3b} and  \ref{fig3c} illustrate the progression of the relative error and regularization parameter at each iteration, respectively, and Fig.~\ref{fig3d} shows the UPRE function at the final iteration. These figures are presented for comparison with the results obtained using the \texttt{Hybrid-RSVD} algorithm. 

For the same gravity problem the results using the \texttt{Hybrid-RSVD} algorithm  for very small values of $q$, $q<100$, are not acceptable.  With increasing $q$ the solution improves  until at $q=m$ the solution matches the \texttt{Hybrid-FSVD} solution. For the reported choices of $q$ all the solutions terminate prior to $\Kmax$, with $K=12$ for $q\ge 150$, and demonstrating that the $\chi^2$ estimate, \eqref{chi2} is satisfied.  We can see that for a suitable value of $q$, the RSVD leads to a solution that is close to that achieved using the \texttt{Hybrid-FSVD}. These conclusions confirm the results in Vatankhah et al. \shortcite{VRA:2018b}; the \texttt{Hybrid-RSVD} algorithm can be used with $q \gtrsim  (m/6)$ for the inversion of gravity data.  We illustrate the results of the inversion using $q=200$ in Fig.~\ref{fig4}.

\begin{table}
\caption{Results of the inversion algorithms applied on gravity data of Fig.~\ref{fig2a}. }\label{gravitytab}
\begin{tabular}{c  c  c  c  c c c c c}
\hline
Method & $q$  &$\alpha^{(1)}$& $\alpha^{(K)}$& $RE^{(K)}$ & $RG^{(K)}$& $K$ & $\chi^2$  & Time (s) \\ \hline
\texttt{Hybrid-FSVD} & $-$  & $61712$ & $32.2$ & $0.7232$ & $-$ &$12$ & $811.8$ & $31.5$ \\ \hline
\texttt{Hybrid-RSVD} & $100$  & $24074$ & $44.5$ & $0.7550$ & $0.0521$ & $14$ & $935.8$ & $37.2$ \\
              & $150$  & $29524$ & $37.4$ & $0.7441$ & $0.0497$ & $12$ & $928.7$ & $32.7$ \\
              & $200$  & $33726$ & $34.5$ & $0.7221$ & $0.0448$ & $12$ & $905.6$ & $32.3$ \\
              & $300$  & $40360$ & $33.4$ & $0.7153$ & $0.0257$ & $12$ & $901.4$ & $33.0$ \\
              & $500$  & $49504$ & $29.5$ & $0.7077$ & $0.0143$ & $12$ & $801.9$ & $35.0$ \\
              & $700$  & $56105$ & $31.1$ & $0.7061$ & $0.0082$ & $12$ & $908.1$ & $35.9$ \\
              & $900$  & $61712$ & $32.2$ & $0.7232$ & $2.8510e^{-14}$ & $12$ & $811.8$ & $37.9$ \\ \hline
\end{tabular}
\end{table}

The results presented for the inversion of the magnetic  data using the \texttt{Hybrid-FSVD} show that the iteration terminates after $8$ iterations, and as indicated in Fig.~\ref{fig5a}  the reconstructed model is in reasonable agreement with the original model. Again a focused image of the subsurface is obtained, but the overestimation of the depth of the left dike is greater as compared with the results in Fig.~\ref{fig3}. Correspondingly, the relative error is larger than that achieved in the inversion of the gravity data. 

\begin{figure*} 
\subfigure{\label{fig7a}\includegraphics[width=.45\textwidth]{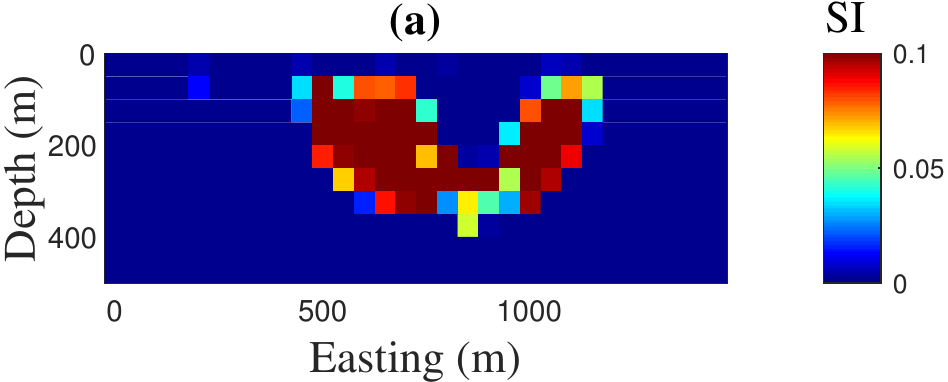}}
\subfigure{\label{fig7b}\includegraphics[width=.45\textwidth]{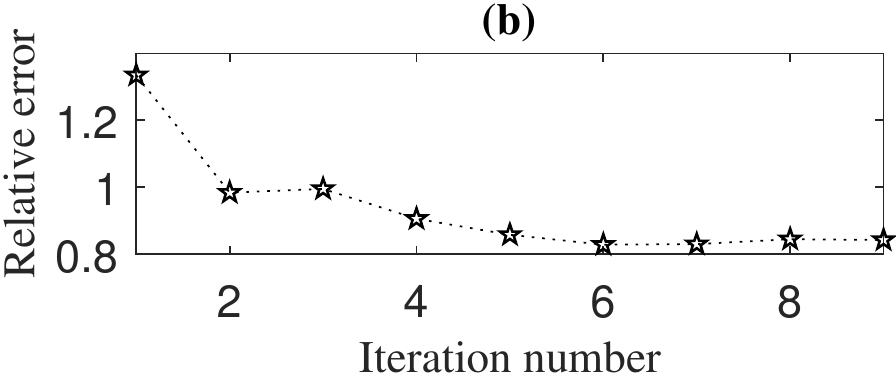}}
\subfigure{\label{fig7c}\includegraphics[width=.45\textwidth]{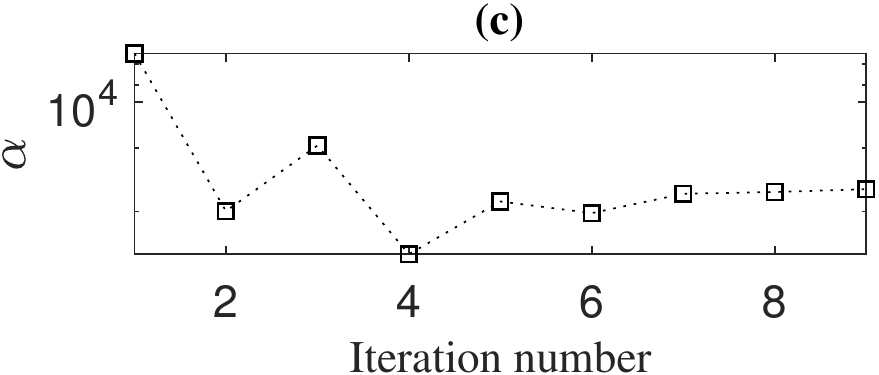}}
\subfigure{\label{fig7d}\includegraphics[width=.45\textwidth]{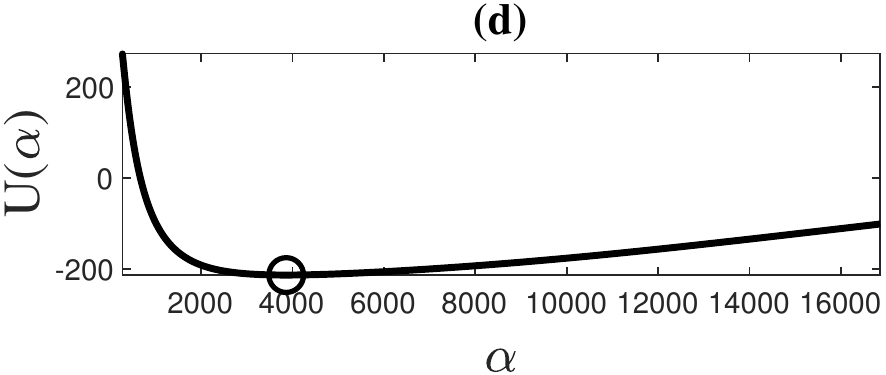}}
\caption{\texttt{Hybrid-RSVD} results using target rank $q=500$ for the inversion of magnetic data given in Fig.~\ref{fig2b}. (a) Cross-section of reconstructed model; (b) The progression of relative error $RE^{(k)}$ with iteration $k$; (c) The progression of regularization parameter $\alpha^{(k)}$ with iteration $k$; (d) The UPRE function at the final iteration.} \label{fig7}
\end{figure*}

For the same magnetic  problem the results using the \texttt{Hybrid-RSVD} algorithm  are not acceptable with reasonable choices for $q$, indeed the algorithm does not terminate prior to $K=\Kmax$ for $q<500$, and the cost therefore increases significantly.   As compared to the inversion of gravity data, larger values of $q$, are required to yield acceptable results. Observe, for example, that  $q=200$ is not a suitable choice because the relative error of the reconstructed model is unacceptable, and the predicted data does not satisfy the observed data at the given noise level; \eqref{chi2} is not satisfied for $k\le 50$. Using $q=500$, the reconstructed model  relative error is reduced and the  inversion terminates at $9$ iterations with an acceptable $\chi^2$ value. We deduce that for the inversion of the magnetic data it is necessary to take $q$ larger than we would use for the inversion of the gravity data, as is indicated by the relatively larger estimates for the rank-$q$ error, $RG^{(K)}$ in Tables~\ref{gravitytab} and \ref{magnetictab}, respectively. To demonstrate the impact of the choice of $q$ we show the results of the inversion for 
 $q=200$ and $q=500$, in Figs.~\ref{fig6} and \ref{fig7}, respectively.

\begin{table}
\caption{Results of the inversion algorithms applied on magnetic data of Fig.~\ref{fig2b}.}\label{magnetictab}
\begin{tabular}{c  c  c  c  c c c c c}
\hline
Method & $q$  &$\alpha^{(1)}$& $\alpha^{(K)}$& $RE^{(K)}$ & $RG^{(K)}$& $K$ & $\chi^2$  & Time (s) \\ \hline
\texttt{Hybrid-FSVD} & $-$    & $21086$ & $4106.4$ & $0.8454$ & $-$ & $8$ & $898.6$  & $20.7$ \\ \hline
\texttt{Hybrid-RSVD} & $100$  & $9866$ & $4372.3$ & $1.0651$ & $0.3735$ & $50$ & $2880.6$  & $125.0$ \\
              & $150$  & $11247$ & $3089.6$ & $0.9904$ & $0.4143$ & $50$ & $2592.5$  & $127.9$ \\
              & $200$  & $12415$ & $3018.0$ & $0.9050$ & $0.3701$ & $50$ & $1673.7$  & $134.9$ \\
              & $300$  & $14371$ & $3030.9$ & $0.8606$ & $0.2948$ & $50$ & $1120.5$  & $135.0$ \\
              & $500$  & $17011$ & $3861.2$ & $0.8426$ & $0.0453$ & $9$ & $917.6$  & $26.9$ \\
              & $700$  & $19071$ & $3854.1$ & $0.8470$ & $0.0268$ & $8$ & $918.2$  & $24.8$ \\
              & $900$  & $21086$ & $4106.4$ & $0.8454$ & $2.8165e^{-14}$ & $8$ & $898.6$  & $26.1$ \\ \hline
\end{tabular}
\end{table}

\subsubsection{The Spectral Properties}\label{sec:spectrum}
\begin{figure*} 
\subfigure{\label{fig8a}\includegraphics[width=.45\textwidth]{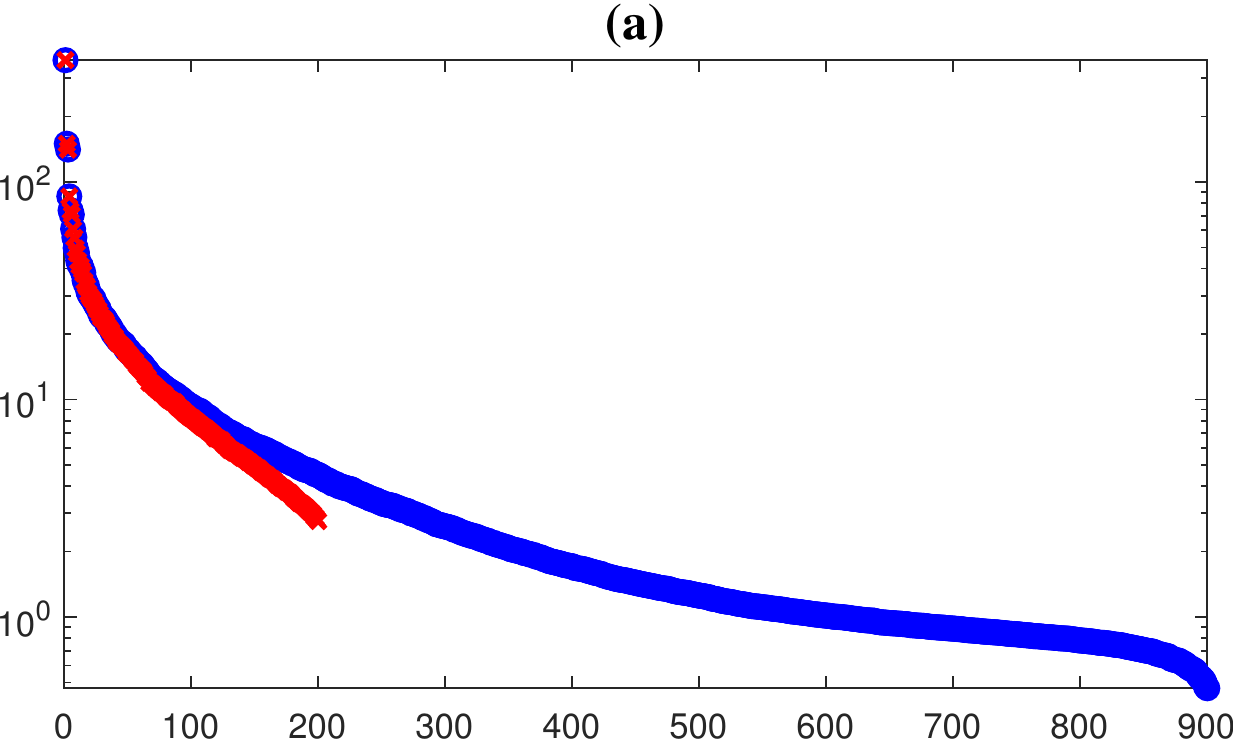}}
\subfigure{\label{fig8b}\includegraphics[width=.45\textwidth]{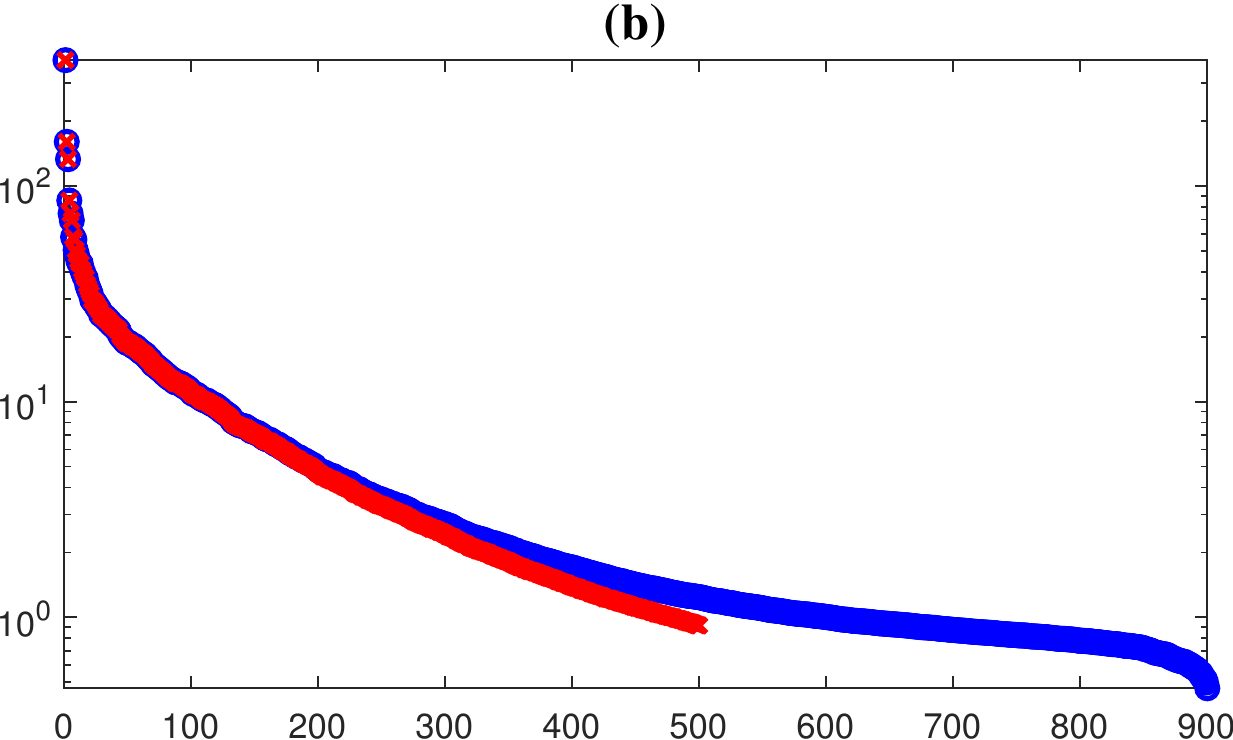}}
\caption{The singular values $\sigma_i^{(k)}$ and $(\sigma_i^{(k)})_q$ for the matrices $\tildetildeG^{(k)}$ (blue circles) and $\tildetildeG^{(k)}_q$ (red crosses), respectively, for the gravity problem. (a) For $q=200$ at iteration $k=8$; (b) For $q=500$ at iteration $k=8$.} \label{fig8}
\end{figure*}

\begin{figure*} 
\subfigure{\label{fig9a}\includegraphics[width=.45\textwidth]{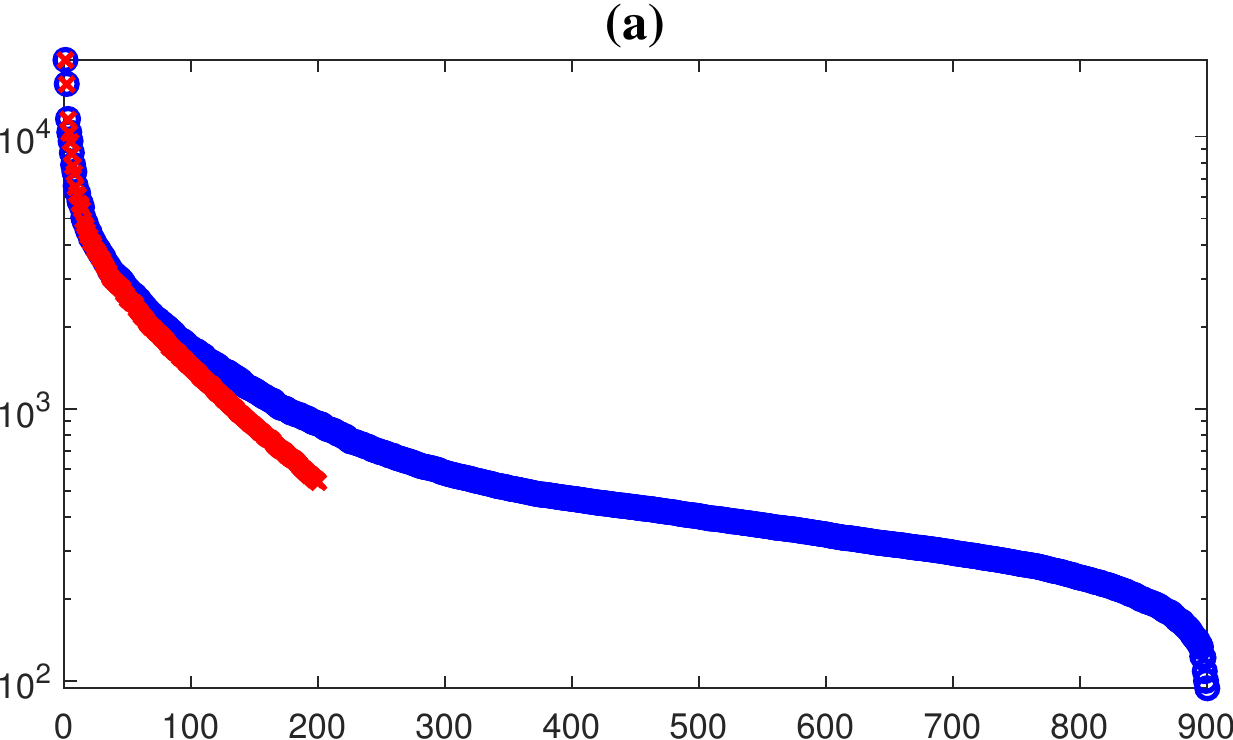}}
\subfigure{\label{fig9b}\includegraphics[width=.45\textwidth]{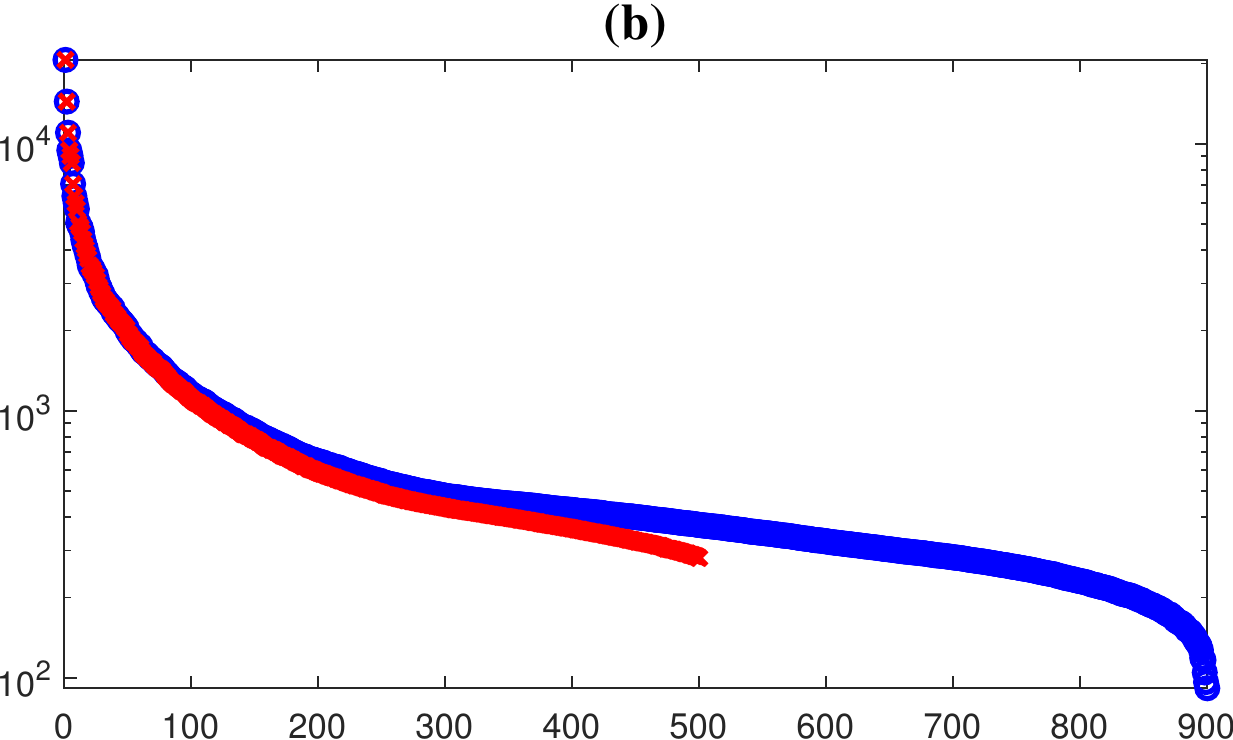}}
\caption {The singular values $\sigma_i^{(k)}$ and $(\sigma_i^{(k)})_q$ for the matrices $\tildetildeG^{(k)}$ (blue circles) and $\tildetildeG^{(k)}_q$ (red crosses), respectively, for the magnetic problem. (a) For $q=200$ at iteration $k=8$ (b) For $q=500$ at iteration $k=8$.} \label{fig9}
\end{figure*}

We now compare the singular values $\sigma_i^{(k)}$ and $(\sigma_i^{(k)})_q$ for the matrices $\tildetildeG^{(k)}$  and $\tildetildeG^{(k)}_q$, respectively, for the gravity and the magnetic problems in Figs.~\ref{fig8} and \ref{fig9}, respectively.  
In both figures, we show the values for $q=200$ and $q=500$ at  iteration $k=8$. It is immediate from these plots that the \texttt{Hybrid-RSVD} algorithm  does not capture the dominant spectrum of the original matrix for the magnetic problem as closely as is the case for the gravity problem. This clarifies why the magnetic inversion requires larger values of $q$  in order for the inversion to  converge. It is also evident that the singular values of both gravity and magnetic problems decay rather slowly, after an initial fast decay. Generally, the RSVD algorithm is more efficient when applied to matrices with rapidly decaying singular values,  as can be seen from the error estimate \eqref{expectederror}. 

\subsubsection{Power Iterations}\label{poweriteration}
Now as discussed in Section~\ref{RSVD} the error estimate \eqref{expectederror} will decrease with increasing $s$. Having seen that the lack of power iterations leads to lack of convergence for the inversion of the magnetic data unless $q$ is taken relatively large, $q \gtrsim m/2$, as compared to just $q\gtrsim m/6$ for the gravity problem, we investigate the power iteration step given by Algorithm~\ref{SubPowerAlgorithm} to improve the column space approximation of  $\tildetildeG^{T}$. We therefore repeat the simulations for the data of the two dike problem illustrated in Figs.~\ref{fig2a}-\ref{fig2b} but employing a power iteration with $s=1$. The results are presented in Tables~\ref{gravitytabpower} and \ref{magnetictabpower} for gravity and magnetic data, respectively, and indicate improvements as compared to the results presented without power iterations. For both problems the number of iterations $K$ is generally reduced, once $q$ is large enough, and the error is generally decreased for a result that used the same number of iterations $K$ with and without power iterations. Moreover, the results are achieved without a large increase in computational cost, where the algorithm converged both with and without power iterations. But the major impact is that it is possible to take a much smaller $q$ to obtain convergence of the inversion of the magnetic problem with reasonable computational cost. In both cases it is sufficient to now take $q=200$ to obtain converged solutions for a relatively small $K$, $11$ and $12$, respectively.

\begin{table}
\caption{Results of the \texttt{Hybrid-RSVD}  algorithm via power iterations applied on gravity data of Fig.~\ref{fig2a}. }\label{gravitytabpower}
\begin{tabular}{c  c  c  c  c c c c}
\hline
$q$  &$\alpha^{(1)}$& $\alpha^{(K)}$& $RE^{(K)}$ & $RG^{(K)}$& $K$ & $\chi^2$  & Time (s) \\ \hline
$100$  & $20629$ & $54.3$ & $0.7018$ & $0.0177$ & $11$ & $913.4$ & $30.4$ \\
$150$  & $25862$ & $45.5$ & $0.7288$ & $0.0152$ & $11$ & $909.6$ & $29.5$ \\
$200$  & $30218$ & $40.8$ & $0.7185$ & $0.0129$ & $11$ & $904.7$ & $30.9$ \\
$300$  & $37279$ & $34.8$ & $0.7188$ & $0.0101$ & $11$ & $898.1$ & $31.8$ \\
$500$  & $47447$ & $29.6$ & $0.7099$ & $0.0063$ & $12$ & $851.0$ & $38.0$ \\
$700$  & $54892$ & $29.0$ & $0.7093$ & $0.0047$ & $12$ & $893.6$ & $39.6$ \\
$900$  & $61712$ & $32.2$ & $0.7232$ & $1.5534e^{-15}$ & $12$ & $811.8$ & $41.3$ \\ \hline
\end{tabular}
\end{table}
 
 \begin{table}
\caption{Results of the \texttt{Hybrid-RSVD} algorithm via power iterations applied on magnetic data of Fig.~\ref{fig2b}. }\label{magnetictabpower}
\begin{tabular}{c  c  c  c  c c c c}
\hline
$q$  &$\alpha^{(1)}$& $\alpha^{(K)}$& $RE^{(K)}$ & $RG^{(K)}$& $K$ & $\chi^2$  & Time (s) \\ \hline
$100$  & $8375$ & $3127.0$ & $0.9187$ & $0.2880$ & $50$ & $1310.6$ & $128.3$ \\
$150$  & $9920$ & $3206.5$ & $0.8464$ & $0.2294$ & $50$ & $955.5$ & $128.7$ \\
$200$  & $11198$ & $5098.5$ & $0.8235$ & $0.0542$ & $12$ & $915.4$ & $33.2$ \\
$300$  & $13235$ & $4447.1$ & $0.8310$ & $0.0323$ & $10$ & $848.9$ & $28.7$ \\
$500$  & $16152$ & $3719.6$ & $0.8421$ & $0.0269$ & $9$ & $814.5$ & $29.0$ \\
$700$  & $18528$ & $4036.8$ & $0.8405$ & $0.0165$ & $8$ & $887.5$ & $26.7$ \\
$900$  & $21086$ & $4106.4$ & $0.8454$ & $2.0710e^{-15}$ & $8$ & $898.6$ & $28.7$ \\ \hline
\end{tabular}
\end{table}

\begin{figure*} 
\subfigure{\label{fig10a}\includegraphics[width=.45\textwidth]{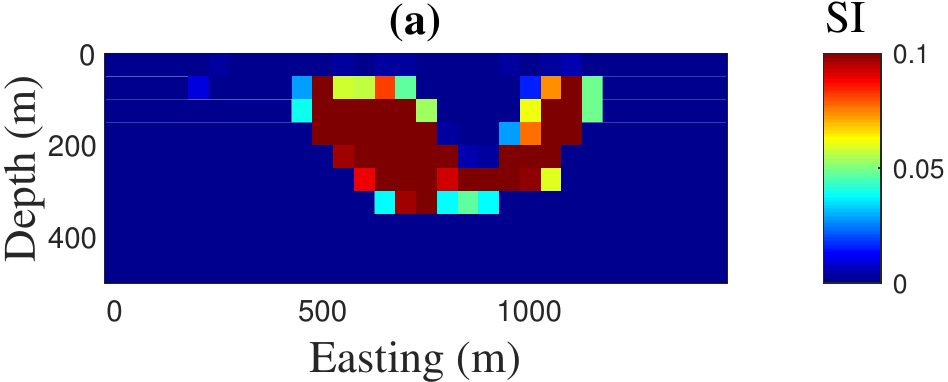}}
\subfigure{\label{fig10b}\includegraphics[width=.45\textwidth]{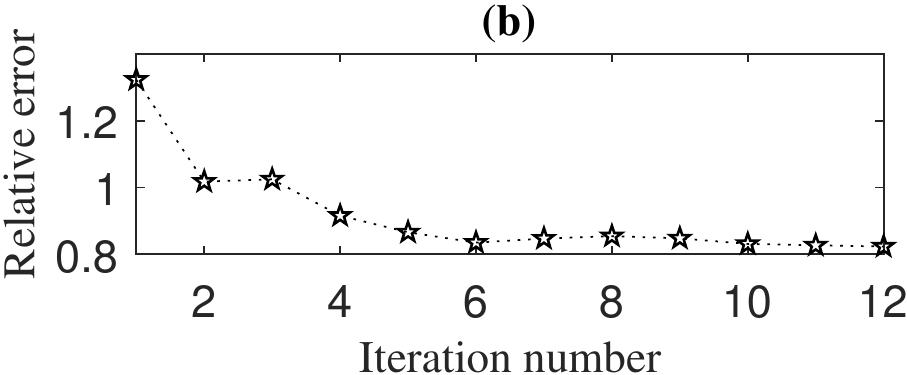}}
\subfigure{\label{fig10c}\includegraphics[width=.45\textwidth]{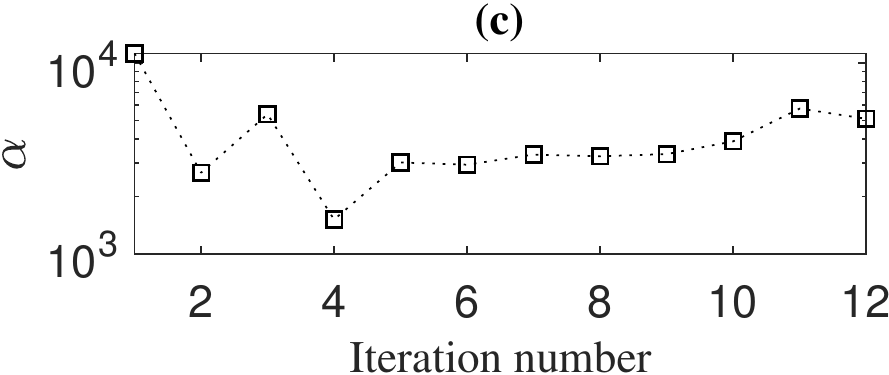}}
\subfigure{\label{fig10d}\includegraphics[width=.45\textwidth]{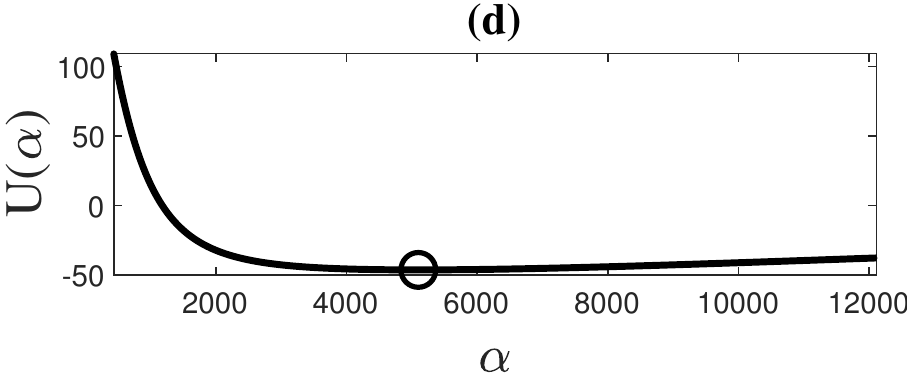}}
\caption {\texttt{Hybrid-RSVD} results using power iterations with $s=1$ and target rank $q=200$ for the inversion of magnetic data given in Fig.~\ref{fig2b}. (a) Cross-section of reconstructed model; (b) The progression of relative error $RE^{(k)}$ with iteration $k$; (c) The progression of regularization parameter $\alpha^{(k)}$ with iteration $k$; (d) The UPRE function at the final iteration.} \label{fig10}
\end{figure*}

\begin{figure*} 
\subfigure{\label{fig11a}\includegraphics[width=.45\textwidth]{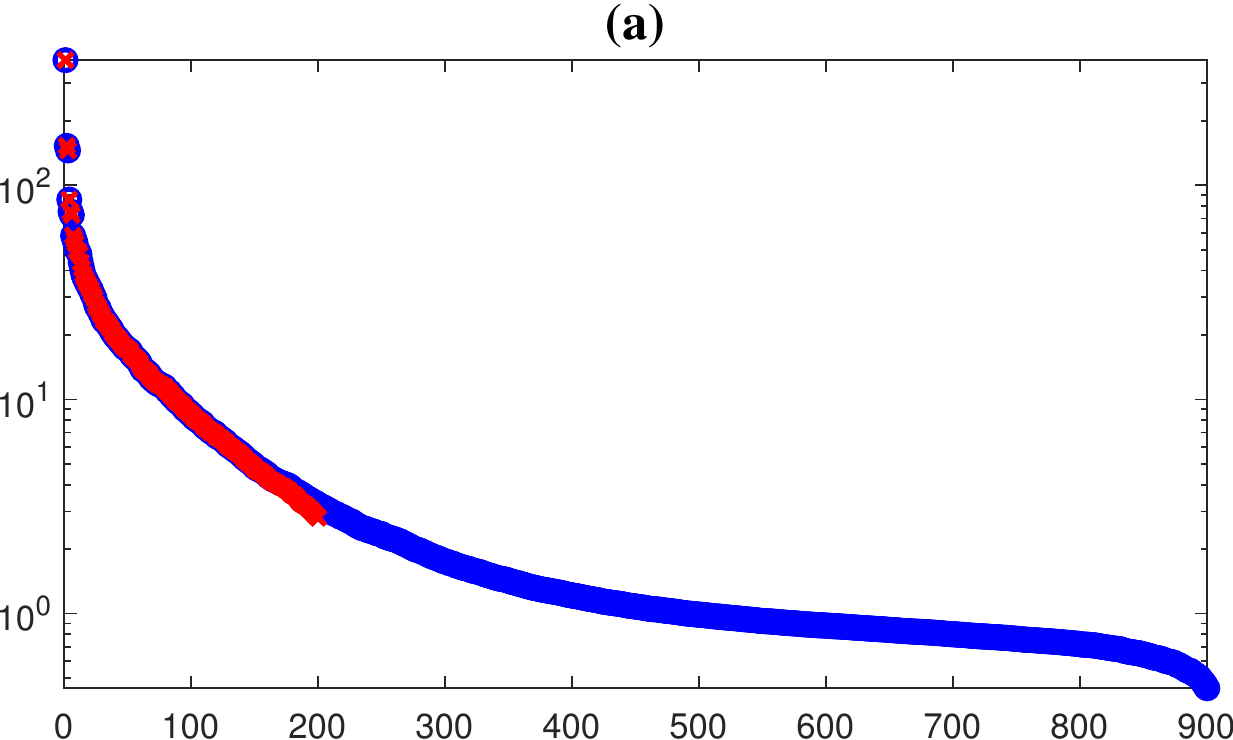}}
\subfigure{\label{fig11b}\includegraphics[width=.45\textwidth]{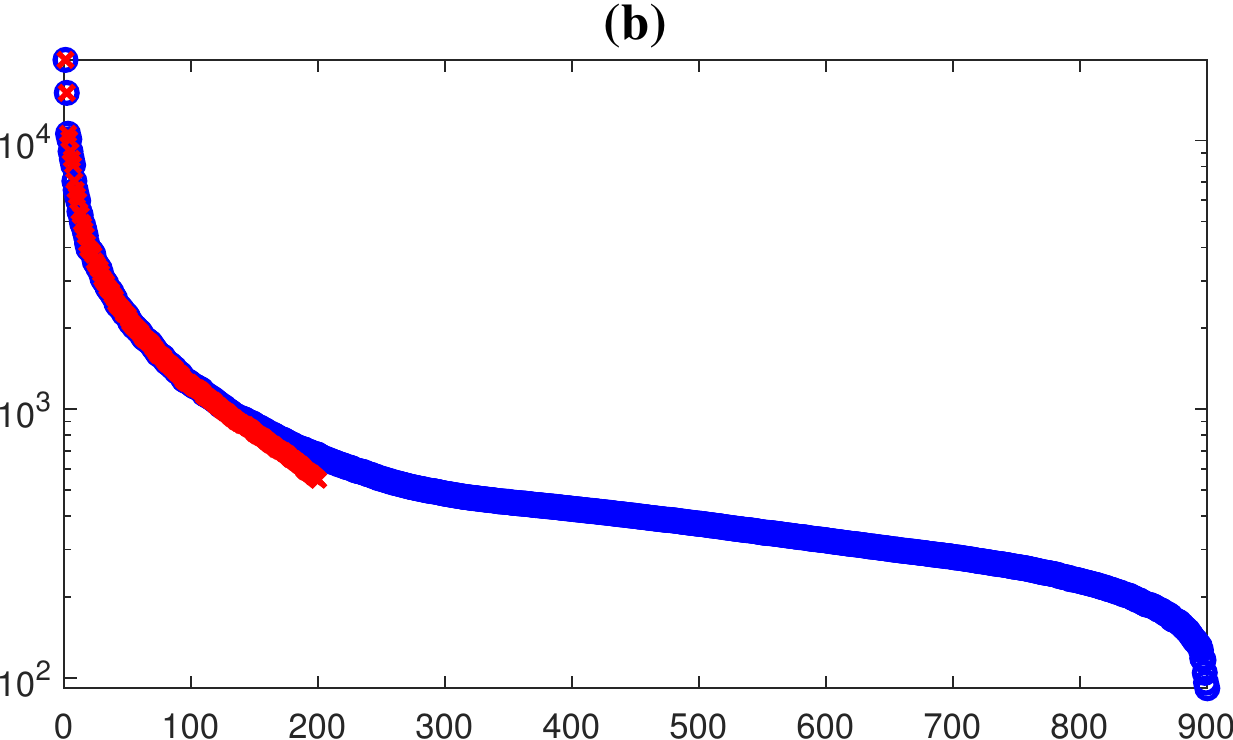}}
\caption {The singular values $\sigma_i^{(k)}$ and $(\sigma_i^{(k)})_q$ for the matrices $\tildetildeG^{(k)}$ (blue circles) and $\tildetildeG^{(k)}_q$ (red crosses), respectively, for $q=200$ and $s=1$. (a) Gravity kernel at iteration $k=8$; (b) Magnetic kernel at iteration $k=8$.}\label{fig11}
\end{figure*}

The results for magnetic data inversion with $q=200$ and $s=1$ are illustrated in Fig.~\ref{fig10} for comparison with Fig.~\ref{fig6} obtained without the power iterations. A noticeable improvement  in the reconstructed model is obtained.  To further show the impact of the power iterations we also show the singular values for the power iterations with $s=1$ and $q=200$ for both gravity and magnetic problems in Figs.~\ref{fig11a}-\ref{fig11b}, comparing with Fig.~\ref{fig8a} and Fig.~\ref{fig9a}, respectively. These plots  demonstrate that the power iterations have indeed improved the accuracy of the estimated singular values.

Naturally these results raise the question as to whether it is better to apply the \texttt{Hybrid-RSVD} algorithm without power iterations and a large choice for $q$ or to use  power iterations and take a smaller $q$. But the main purpose of using Algorithm~\ref{RSVDAlgorithm}  is to make it feasible, in terms of both memory and computational cost, to find accurate solutions of  large scale problems. Indeed the aim is to solve problems which are either too expensive to solve using the \texttt{Hybrid-FSVD} or cannot be solved at all using the \texttt{Hybrid-FSVD}. We discuss this further for a larger problem in Section~\ref{multiplebodies}

\subsection{Model of multiple bodies}\label{multiplebodies}
\begin{figure*} 
\includegraphics[width=.8\textwidth]{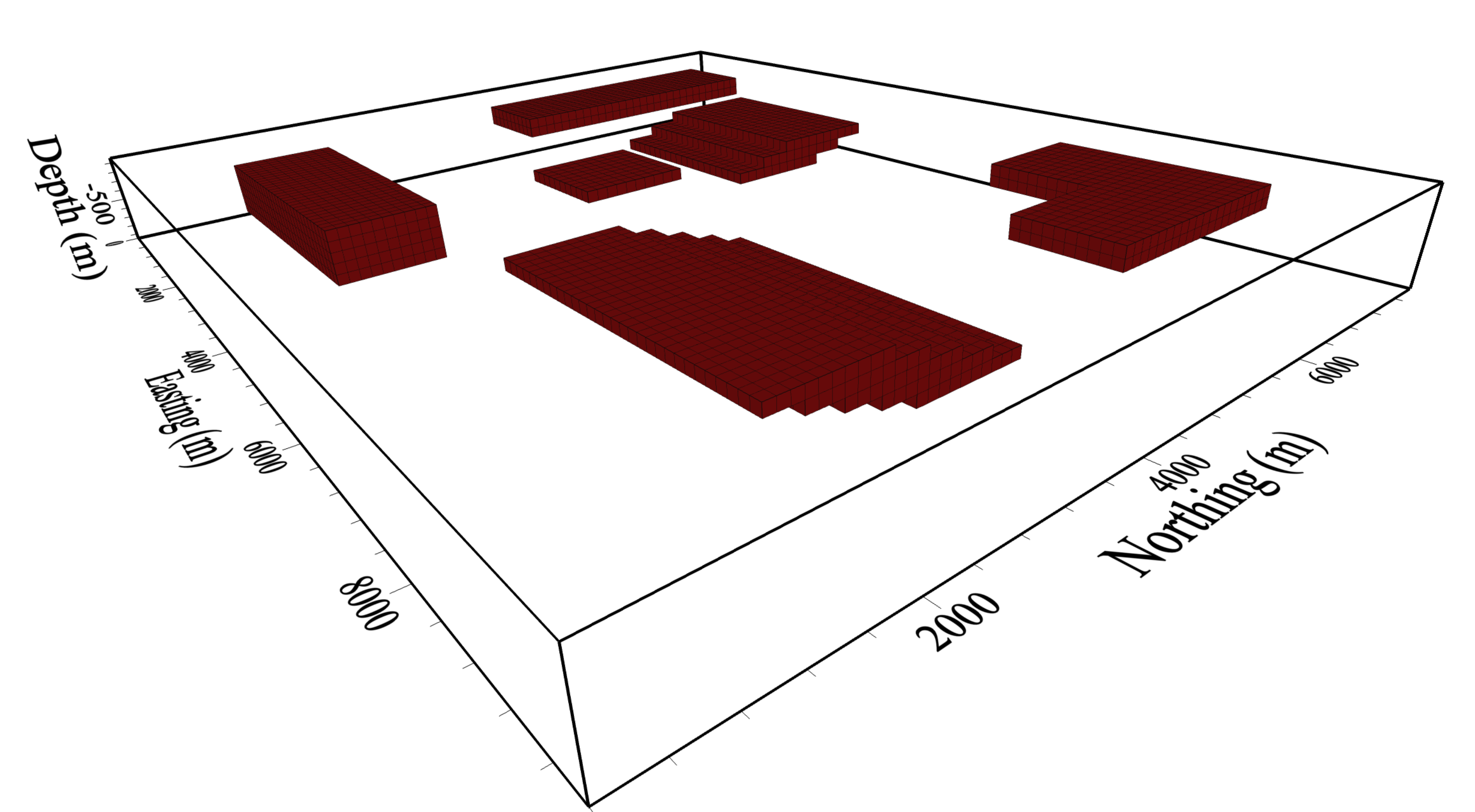}
\caption {Model consisting of  six bodies with different shapes, depths and dimensions.} \label{fig12}
\end{figure*}

\begin{figure*} 
\subfigure{\label{fig13a}\includegraphics[width=.45\textwidth]{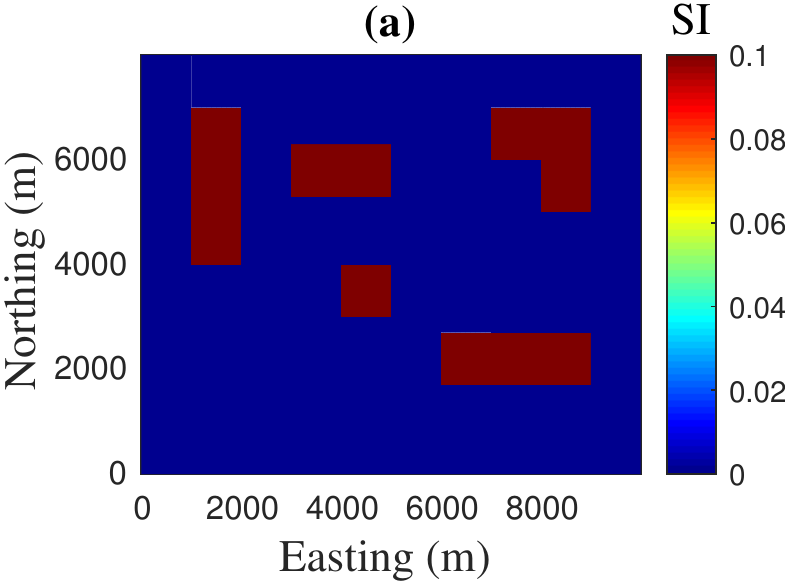}}
\subfigure{\label{fig13b}\includegraphics[width=.45\textwidth]{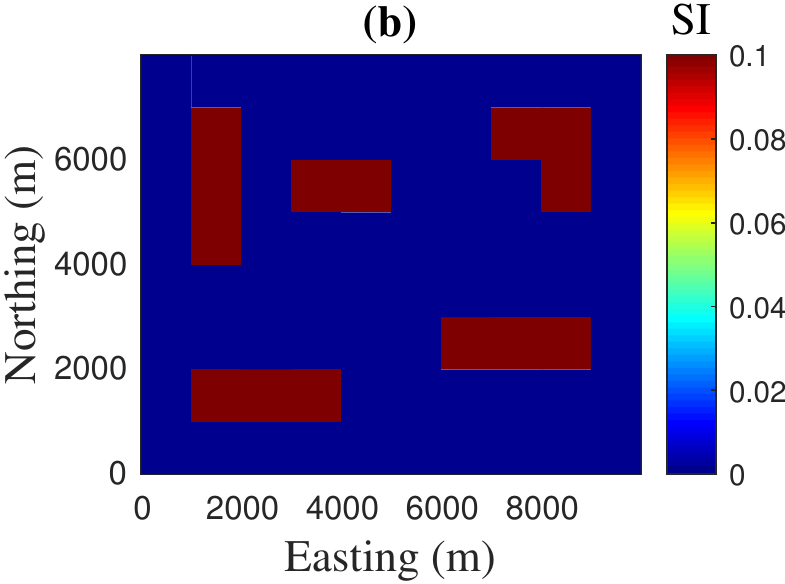}}
\subfigure{\label{fig13c}\includegraphics[width=.45\textwidth]{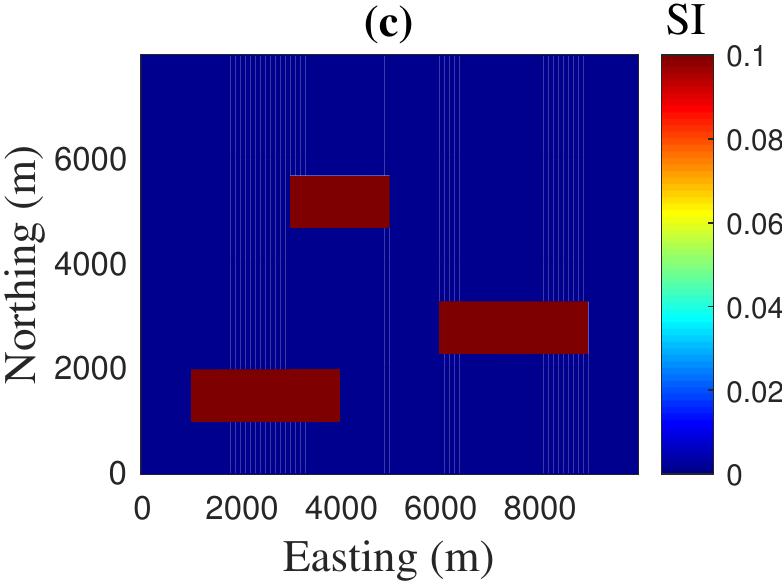}}
\subfigure{\label{fig13d}\includegraphics[width=.45\textwidth]{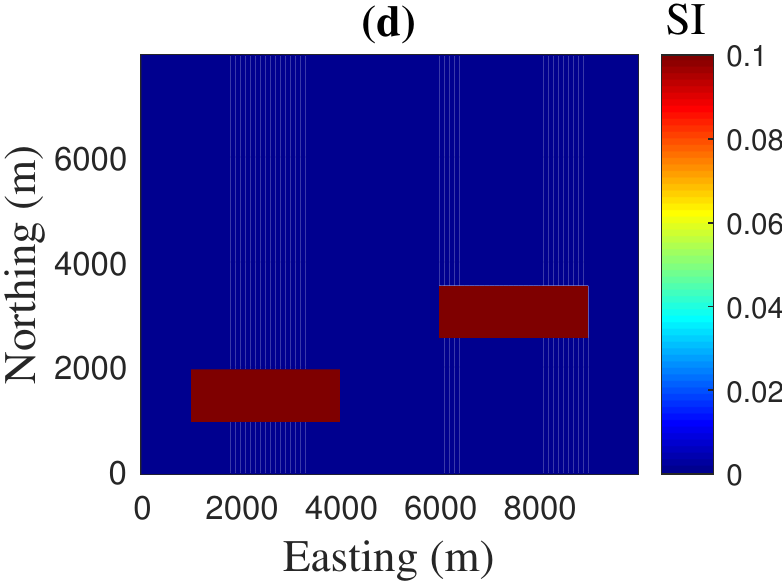}}
\subfigure{\label{fig13e}\includegraphics[width=.45\textwidth]{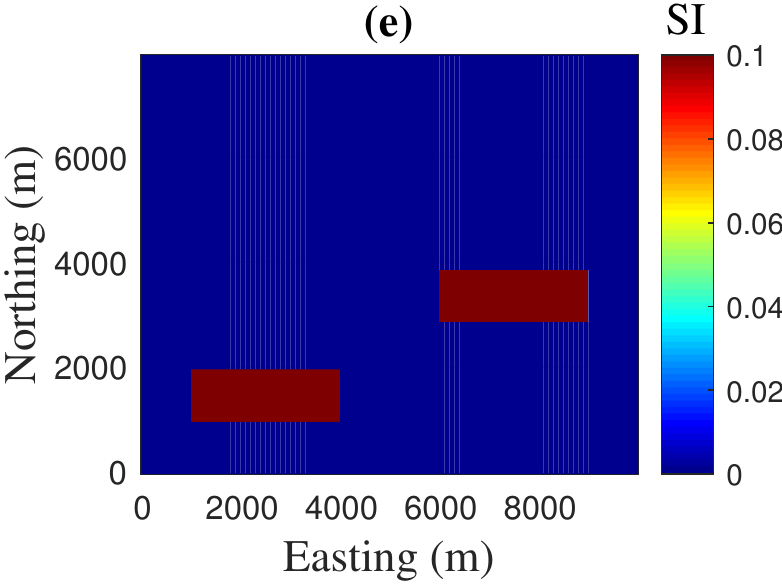}}
\subfigure{\label{fig13f}\includegraphics[width=.45\textwidth]{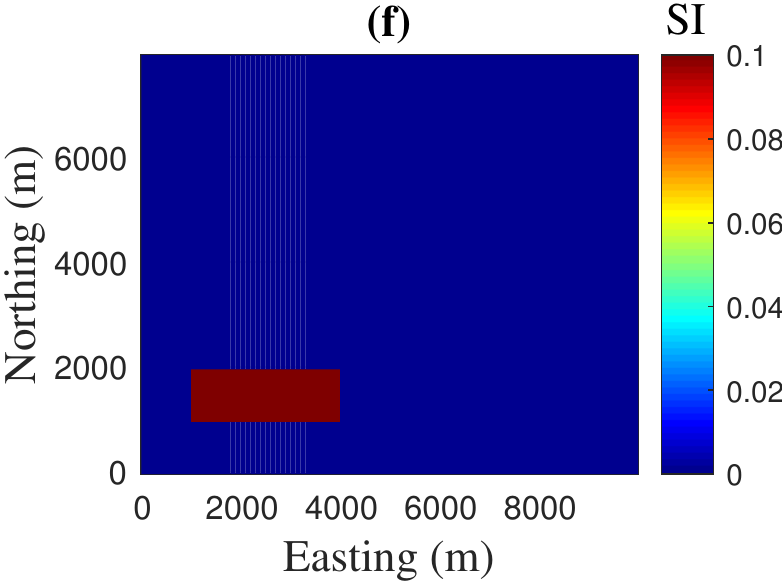}}
\caption {The susceptibility distribution of the model in Fig.~\ref{fig12} is displayed in six plane-sections. The depth of the sections are: (a) $100$~m;  (b) $200$~m; (c) $300$~m; (d) $400$~m; (e) $500$~m; and (f) $600$~m.} \label{fig13}
\end{figure*}

\begin{figure*} 
\subfigure{\label{fig14a}\includegraphics[width=.43\textwidth]{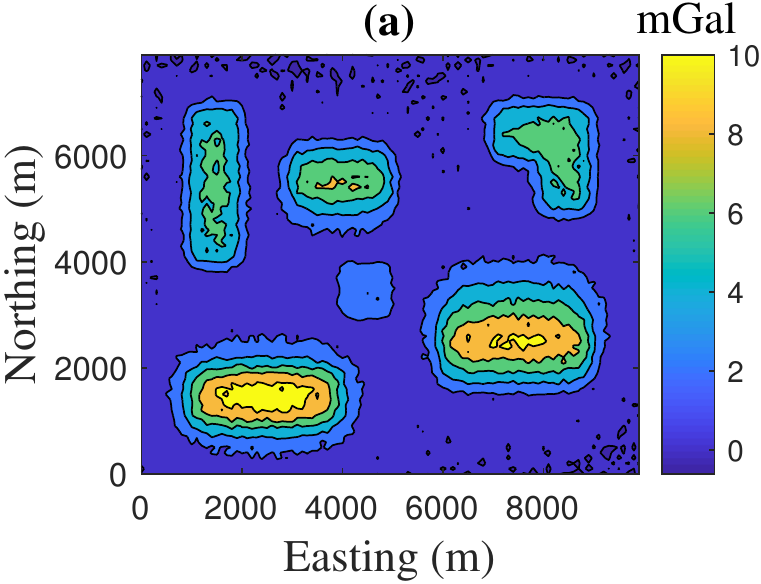}}
\subfigure{\label{fig14b}\includegraphics[width=.45\textwidth]{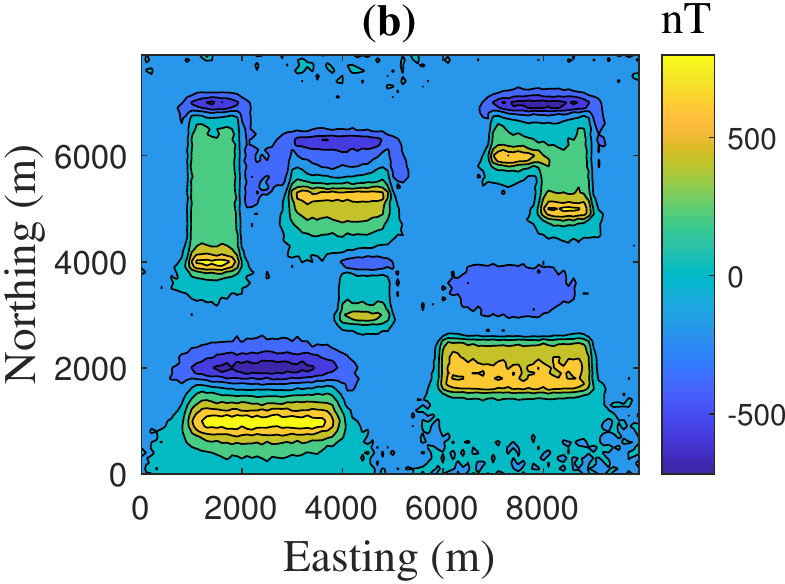}}
\caption {Anomaly produced by the model shown in Fig.~\ref{fig12} and contaminated by Gaussian noise.   (a) Vertical component of the gravity field. The noise parameters are $(\tau_1=0.02, \tau_2=0.02)$; (b) Total magnetic field. Here, $(\tau_1=0.02, \tau_2=0.018)$. The $\mathrm{SNR}$ for  gravity and magnetic data, respectively, are $22.0348$ and $21.8817$. } \label{fig14}
\end{figure*}

We now study the application of the \texttt{Hybrid-RSVD} algorithm for the solution of a larger and more  complex model consisting of six different bodies with different shapes, dimensions and depths, as shown in the perspective view in Fig.~\ref{fig12} and the six plane-sections in Fig.~\ref{fig13}. The data for the problem, the vertical component of the gravity and the total magnetic field, are generated on the surface on a   $100 \times 80$ grid with $100$~m spacing.  The noisy gravity and magnetic data in each case are illustrated in Figs.~\ref{fig14a} and \ref{fig14b}. 

To perform the inversion, the subsurface volume is discretized into $100 \times 80 \times 10$  cubes of size $100$~m in each dimension. The resulting matrix $\tildetildeG$  is of size $8000 \times 80000$ and is too large for efficient use of the \texttt{Hybrid-FSVD}.  The results obtained using the \texttt{Hybrid-RSVD} algorithm with the choices $q=1000$, $1500$, $2500$, and $4200$ are detailed in Tables~\ref{gravitytablarge} and \ref{magnetictablarge}, respectively, both with and without power iterations. 
\begin{table}
\caption{Results of the inversion algorithms applied on the gravity data of Fig.~\ref{fig14a}. }\label{gravitytablarge}
\begin{tabular}{c  c  c  c  c c c c c}
\hline
Method & $q$  &$\alpha^{(1)}$& $\alpha^{(K)}$& $RE^{(K)}$ & $K$ & $\chi^2$  & Time (s) \\ \hline
\texttt{Hybrid-RSVD} & $1000$  & $40276$ & $38.4$ & $0.7389$ &  $21$ & $8108.5$ & $376.8$ \\
                     & $1500$  & $48847$ & $44.6$ & $0.7168$ &  $20$ & $7706.2$ & $524.3$ \\
                     & $2500$  & $61439$ & $39.1$ & $0.7111$ &  $19$ & $7893.3$ & $946.6$ \\
                     & $4200$  & $76008$ & $37.9$ & $0.7013$ &  $20$ & $6943.7$ & $2114.0$ \\ \hline
\texttt{Hybrid-RSVD} & $1000$  & $34324$ & $38.1$ & $0.6927$ &  $21$ & $7915.9$ & $617.0$ \\
 with power iterations        & $1500$  & $43000$ & $36.7$ & $0.6976$ &  $20$ & $7985.6$ & $886.3$ \\
              & $2500$  & $56392$ & $40.0$ & $0.6942$ &  $20$ & $7561.5$ & $1775.4$ \\
              & $4200$  & $72451$ & $38.2$ & $0.6986$ &  $20$ & $7989.2$ & $4178.8$ \\ \hline  
\end{tabular}
\end{table}

\begin{table}
\caption{Results of the inversion algorithms applied on the magnetic data of Fig.~\ref{fig14b}.}\label{magnetictablarge}
\begin{tabular}{c  c  c  c  c c c c c}
\hline
Method & $q$  &$\alpha^{(1)}$& $\alpha^{(K)}$& $RE^{(K)}$ & $K$ & $\chi^2$  & Time (s) \\ \hline
\texttt{Hybrid-RSVD} & $1000$  & $12600$ & $10774.9$ & $1.2110$ &  $50$ & $26292.4$ & $887.9$ \\
                     & $1500$  & $14564$ & $8947.8$ & $1.1720$ &  $50$ & $20789.4$ & $1312.4$ \\
                     & $2500$  & $17376$ & $9253.8$ & $1.1063$ &  $50$ & $11891.4$ & $2520.0$ \\
                     & $4200$  & $20695$ & $8778.1$ & $1.0975$ &  $13$ & $7750.3$ & $1435.9$ \\  \hline                 
\texttt{Hybrid-RSVD} & $1000$  & $10705$ & $11128.1$ & $1.1315$ &  $50$ & $13480.7$ & $1459.1$ \\
with power iterations             & $1500$  & $12744$ & $8992.1$ & $1.0910$ &  $50$ & $9570.6$ & $2263.9$ \\ 
              & $2500$  & $15837$ & $8425.2$ & $1.0947$ &  $14$ & $7863.0$ & $1260.4$ \\
              & $4200$  & $19518$ & $8665.2$ & $1.1080$ &  $12$ & $7299.8$ & $2685.4$ \\ \hline 
\end{tabular}
\end{table}

As with the inversion of the two-dike problem, it is immediate that the inversions for the gravity problem are acceptable, without power iterations, for far smaller $q$ than for the magnetic case. Applying the power iterations reduces the size of $q$ that is required to obtain convergence and excellent results are obtained with just $q=1000$ and a time that is little more than is required for $q=1500$ and no power iterations. This suggests that we may use $q\gtrsim m/8$ with $s=1$ for the power iterations. As for the two-dike problem the magnetic inversion iteration does not converge to the required $\chi^2$ level within $50$ iterations, except when we take $q=4200>m/2$. Further, the relative errors are large and the computational cost is high. Including the power iterations yields convergence at $K=14$ when $q=2500$  and a smaller relative error for an acceptable computational cost. These results suggest that it may be acceptable to take   $q\gtrsim m/4$ in the inversion of magnetic data with the \texttt{Hybrid-RSVD} algorithm combined with $s=1$ power iterations. This choice yields run times for the magnetic inversion that are comparable to those for the gravity inversion.  

\begin{figure*} 
\subfigure{\label{fig15a}\includegraphics[width=.45\textwidth]{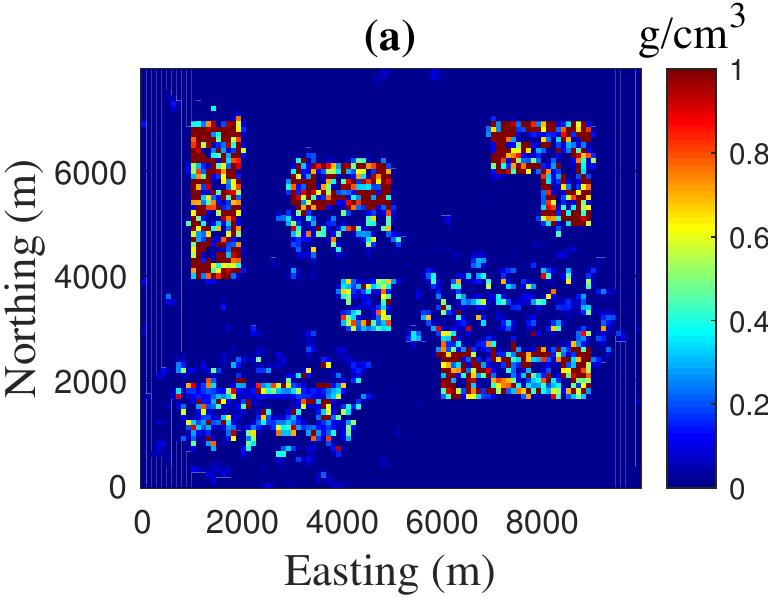}}
\subfigure{\label{fig15b}\includegraphics[width=.45\textwidth]{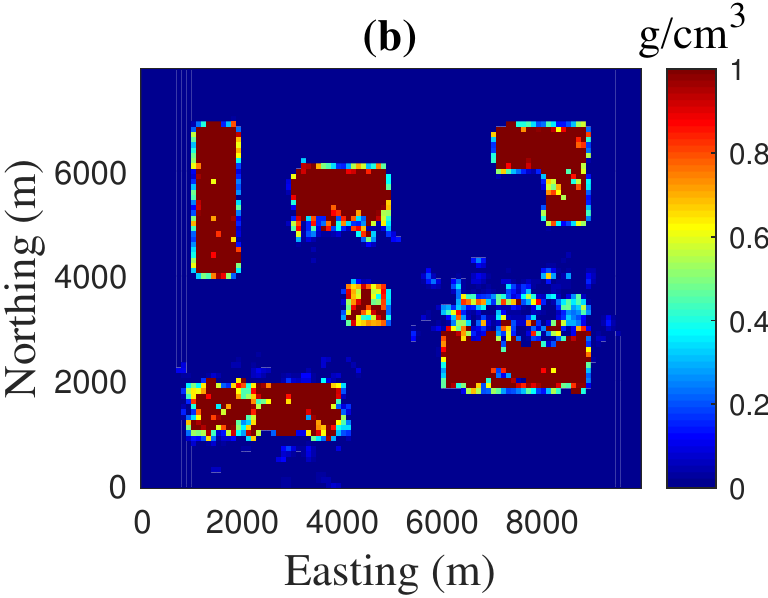}}
\subfigure{\label{fig15c}\includegraphics[width=.45\textwidth]{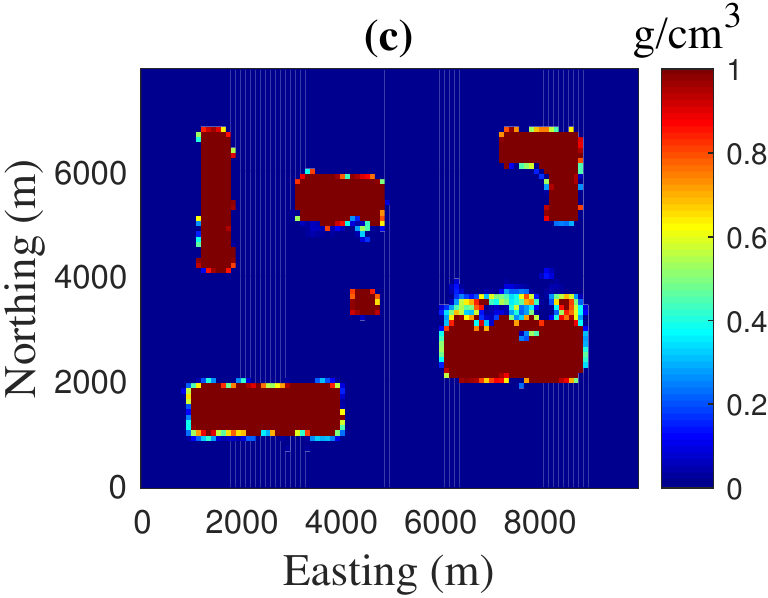}}
\subfigure{\label{fig15d}\includegraphics[width=.45\textwidth]{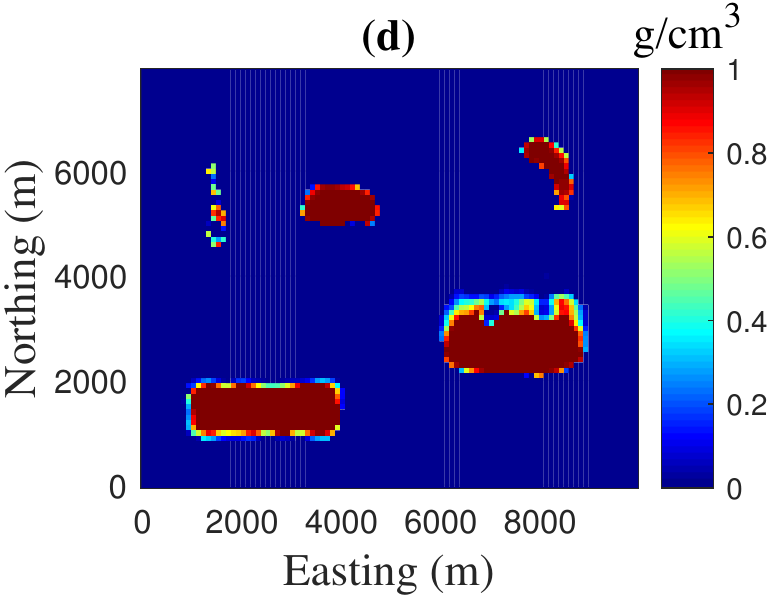}}
\subfigure{\label{fig15e}\includegraphics[width=.45\textwidth]{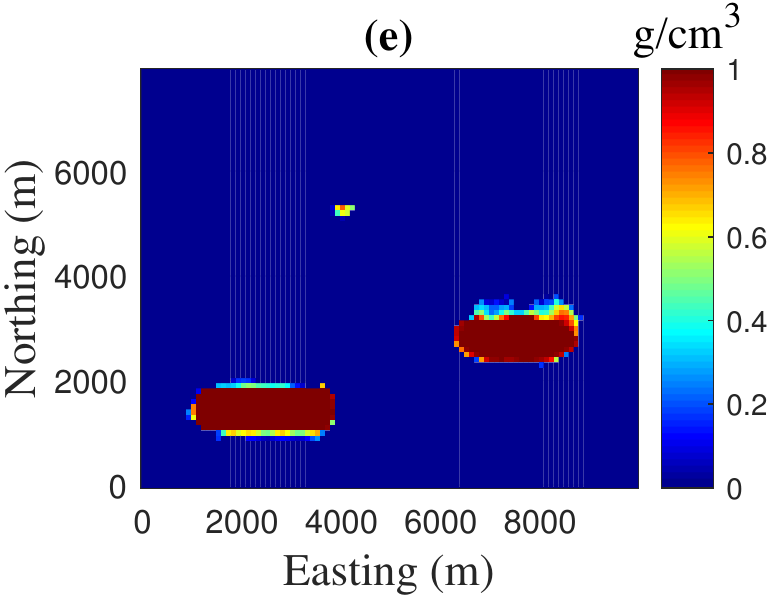}}
\subfigure{\label{fig15f}\includegraphics[width=.45\textwidth]{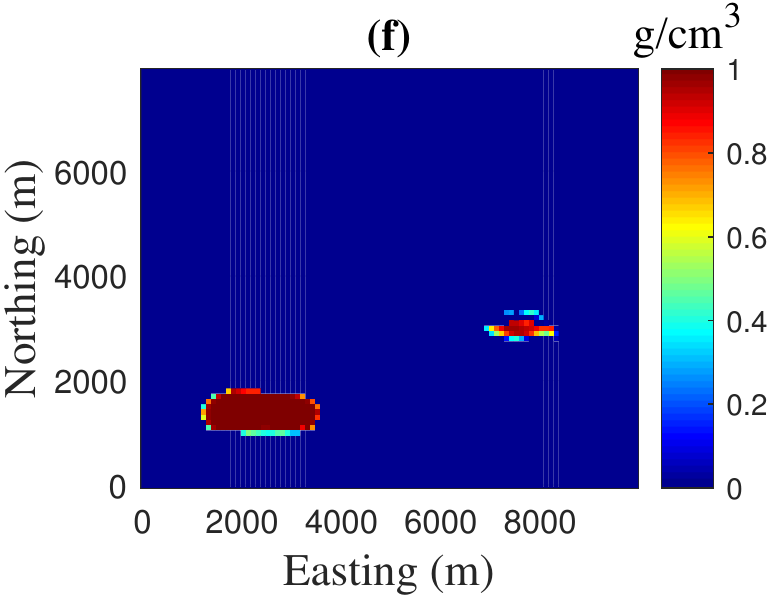}}
\caption {\texttt{Hybrid-RSVD} density results using power iteration with $s=1$ and target rank $q=2500$ for the inversion of gravity data given in Fig.~\ref{fig14a}.
The depth of the sections are: (a) $100$~m;  (b) $200$~m; (c) $300$~m; (d) $400$~m; (e) $500$~m; and (f) $600$~m.} \label{fig15}
\end{figure*}

\begin{figure*} 
\subfigure{\label{fig16a}\includegraphics[width=.45\textwidth]{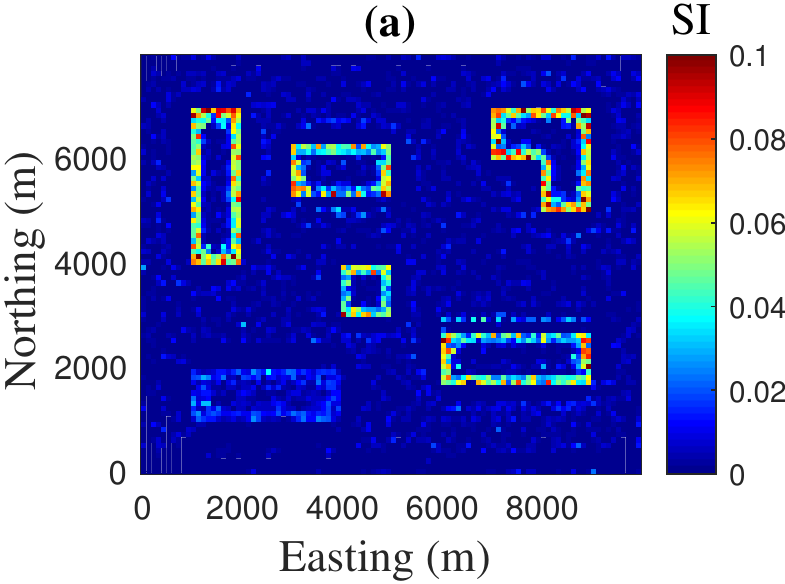}}
\subfigure{\label{fig16b}\includegraphics[width=.45\textwidth]{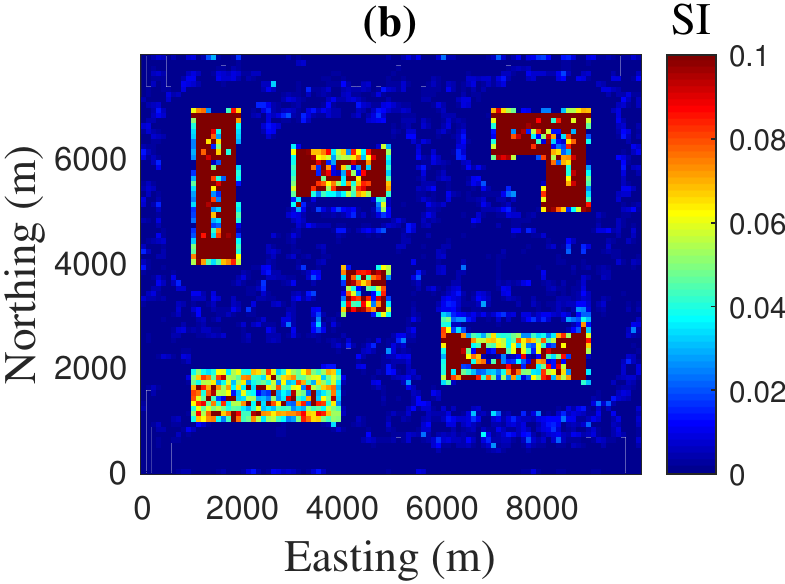}}
\subfigure{\label{fig16c}\includegraphics[width=.45\textwidth]{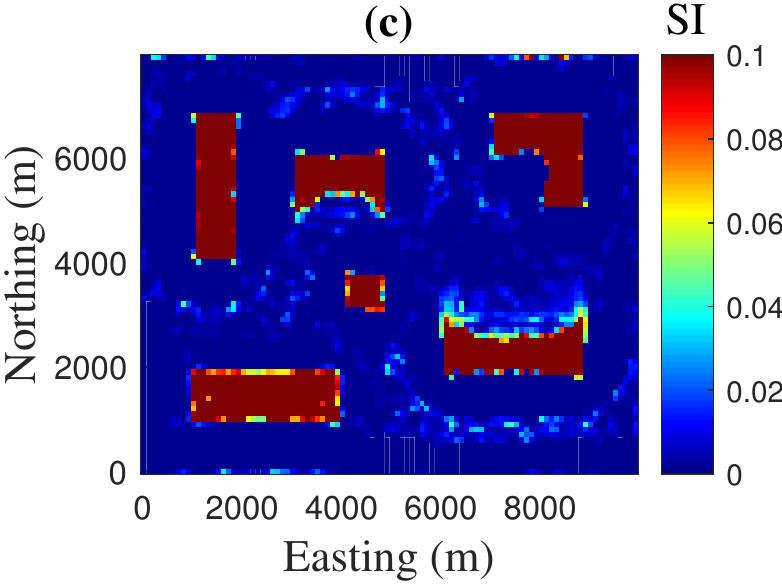}}
\subfigure{\label{fig16d}\includegraphics[width=.45\textwidth]{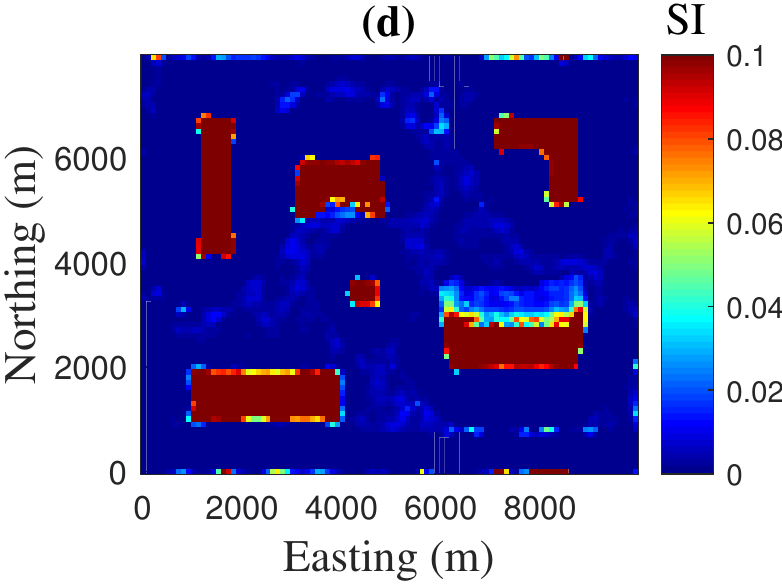}}
\subfigure{\label{fig16e}\includegraphics[width=.45\textwidth]{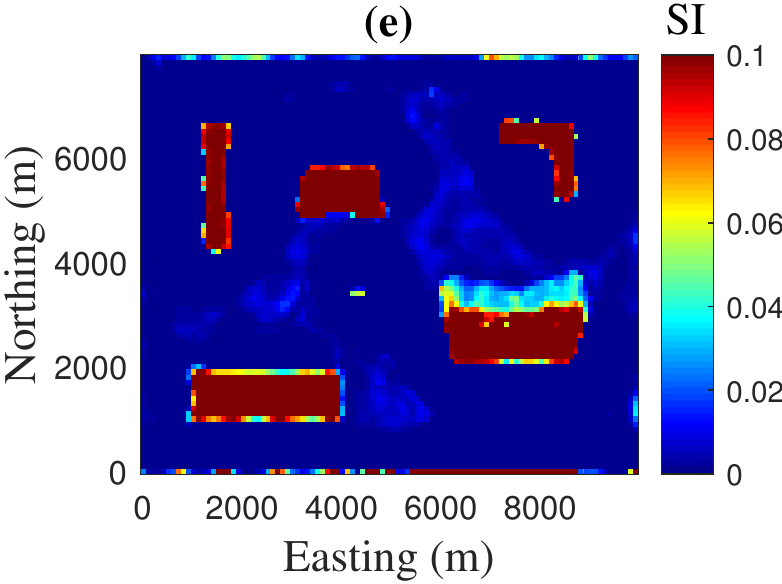}}
\subfigure{\label{fig16f}\includegraphics[width=.45\textwidth]{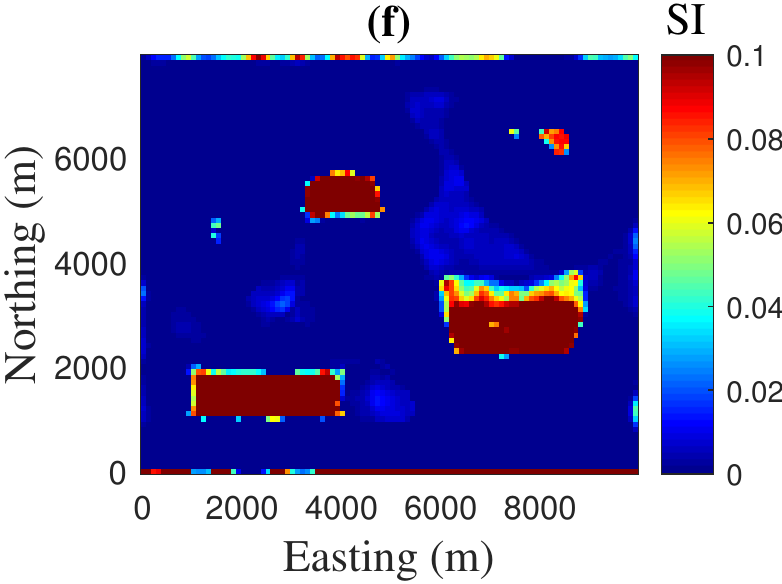}}
\caption {\texttt{Hybrid-RSVD} susceptibility results using power iteration with $s=1$ and target rank $q=2500$ for the inversion of magnetic data given in Fig.~\ref{fig14b}. The depth of the sections are: (a) $100$~m;  (b) $200$~m; (c) $300$~m; (d) $400$~m; (e) $500$~m; and (f) $600$~m.} \label{fig16}
\end{figure*}

\begin{figure*} 
\includegraphics[width=.8\textwidth]{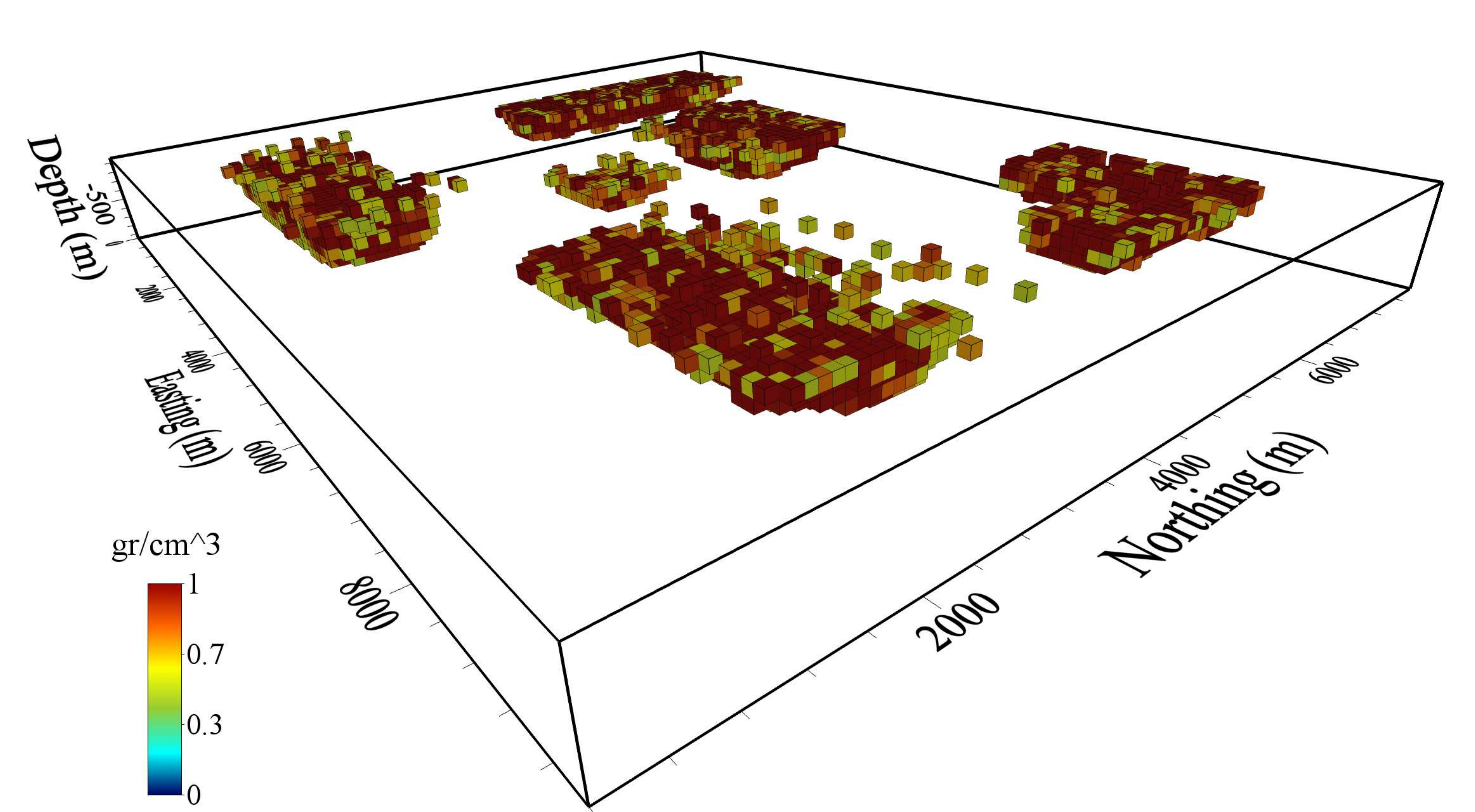}
\caption {$3$-D view of the reconstructed density model shown in Fig.~\ref{fig15}, illustrating cells with  $\rho>0.5$~g~cm$^{-3}$.} \label{fig17}
\end{figure*}

\begin{figure*} 
\includegraphics[width=.8\textwidth]{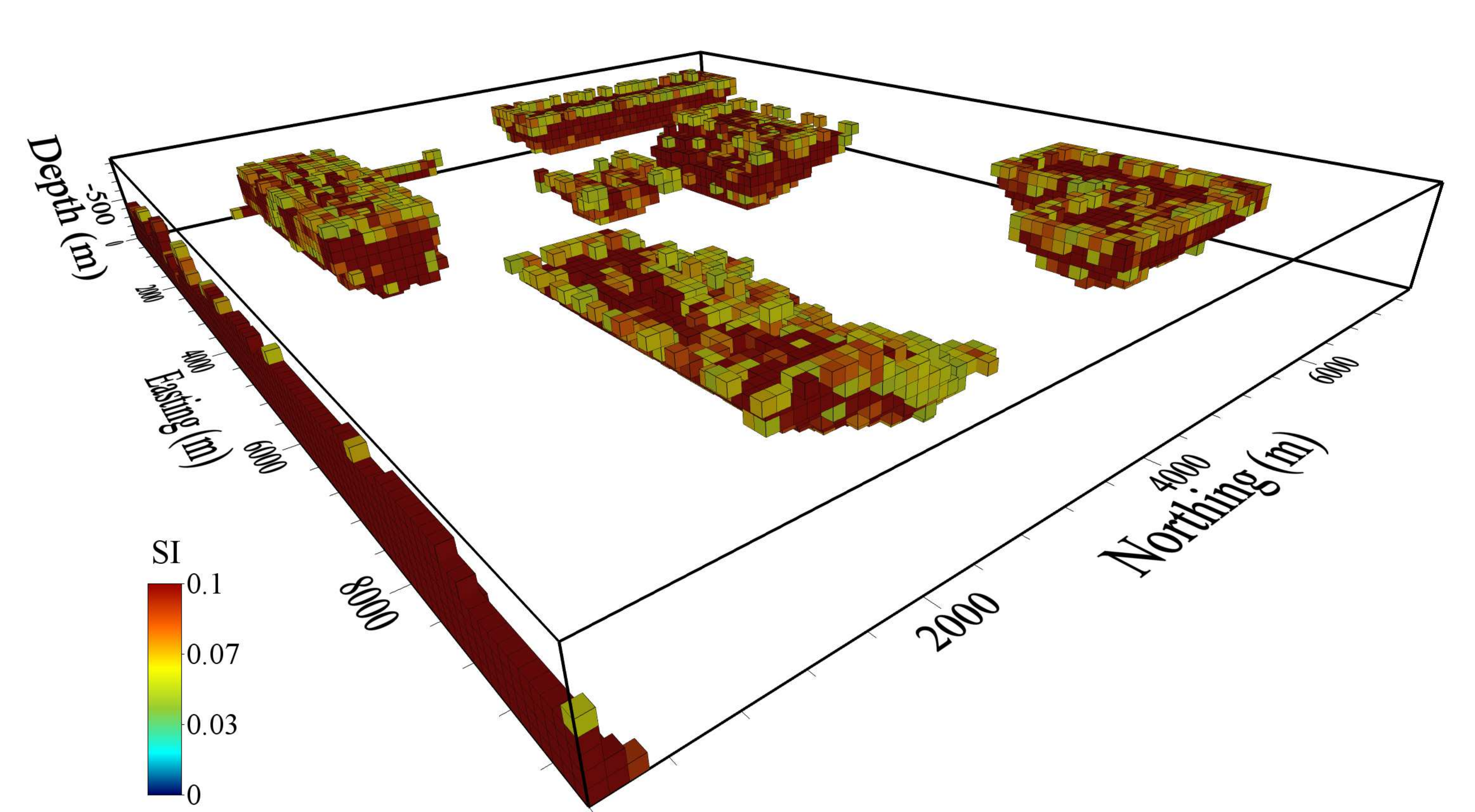}
\caption {$3$-D view of the reconstructed susceptibility model shown in Fig.~\ref{fig16}, illustrating cells with  $\kappa>0.05$ (SI unit).} \label{fig18}
\end{figure*}

The cross-sections for the  inversions using the power iterations and $q=2500$  for the gravity and magnetic data   are given in Fig.~\ref{fig15} and Fig.~\ref{fig16}, respectively. The perspective view of these solutions is also given in Figs.~\ref{fig17} and \ref{fig18}, respectively. We observe that the horizontal borders of the reconstructed models, in both inversions, are in good agreement
with those of the original model, but that additional structures appear at depth. Here, the reconstructed model of the magnetic susceptibility exhibits more artifacts at depth and has a higher relative error than achieved for the gravity reconstruction.   On the other hand, the magnetic structure better illustrates  the dip of both dikes which is significant for accurate geophysical interpretation of the structures. Moreover, these results indicate that  joint interpretation of the individual magnetic and gravity inversions may improve the quality of the final subsurface model.

\section{Real data}\label{real}
\begin{figure*} 
\subfigure{\label{fig19a}\includegraphics[width=.45\textwidth]{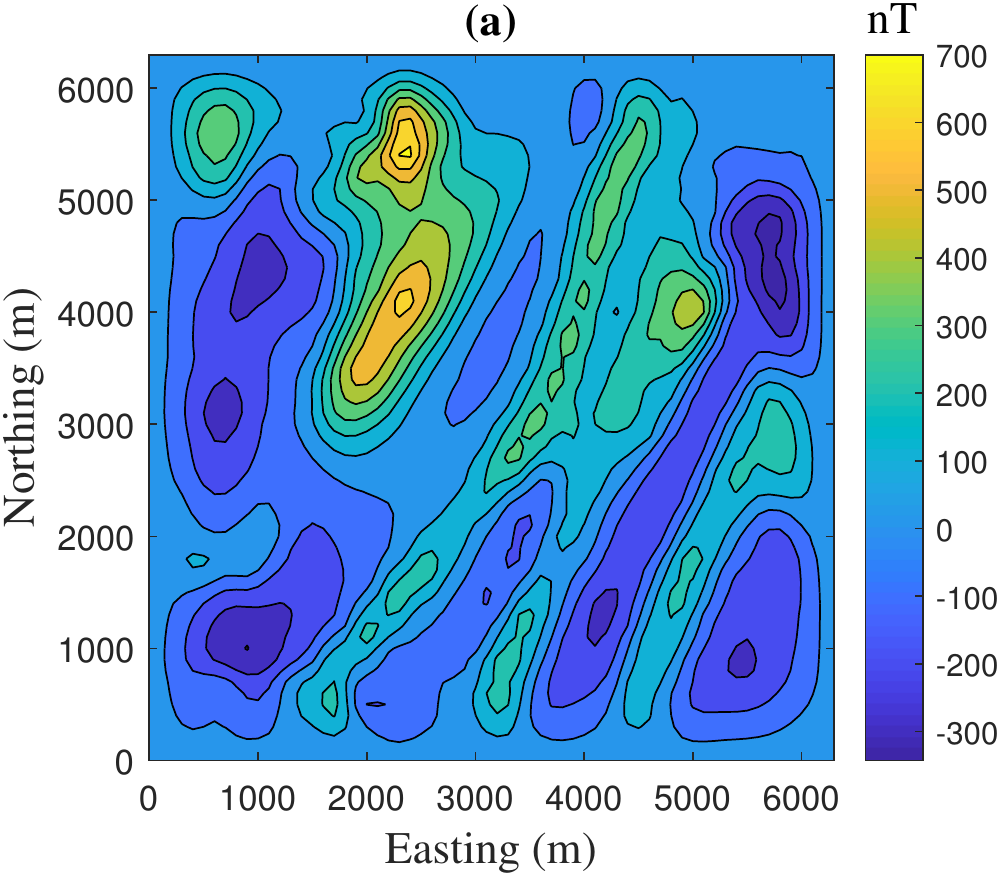}}
\subfigure{\label{fig19b}\includegraphics[width=.45\textwidth]{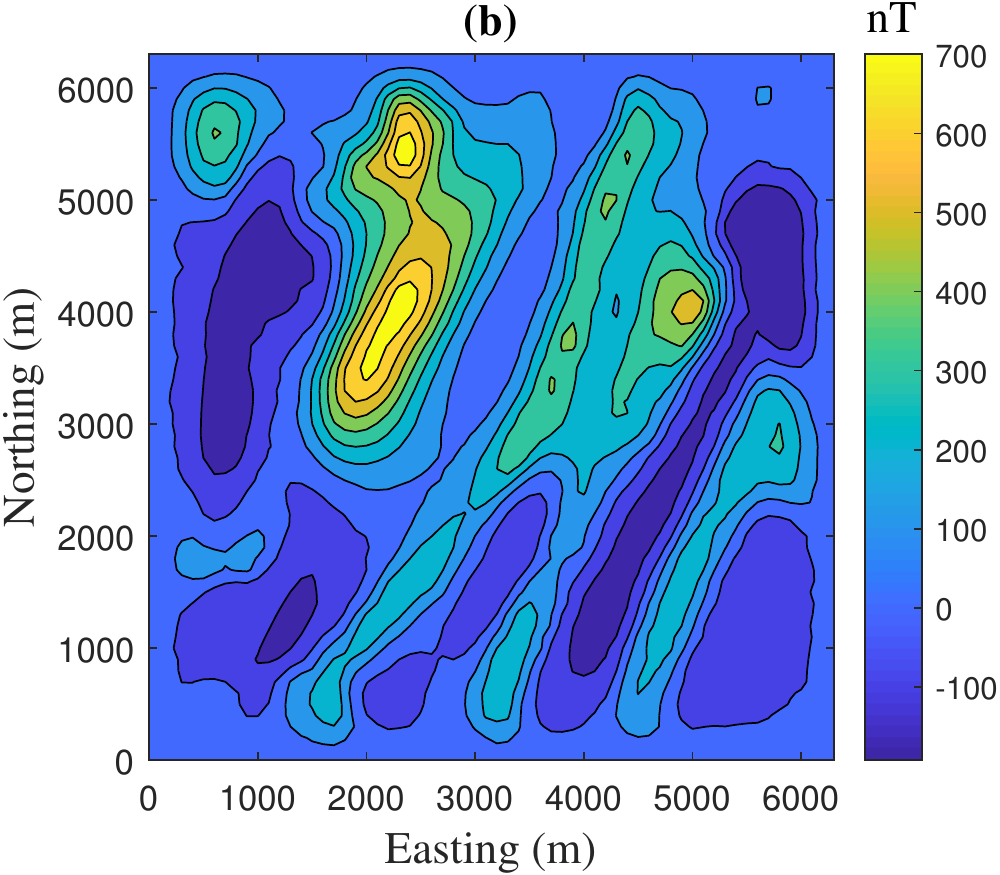}}
\caption { (a) Total magnetic field over a portion of the Wuskwatim
Lake region in Manitoba, Canada; (b) The anomaly produced by reconstructed model in Fig.~\ref{fig20}. } \label{fig19}
\end{figure*}

The \texttt{Hybrid-RSVD} algorithm  with and without power iterations, is now applied for the inversion of total magnetic field data that have been obtained over a portion of the Wuskwatim
Lake region in Manitoba, Canada, Fig.~\ref{fig19a}\footnote{Results showing the application of the methodology for real gravity data were already presented in Vatankhah et al. \shortcite{VRA:2018b}.}. The given area lies within a poorly exposed meta-sedimentary gneiss belt consisting of paragneiss, amphibolite, and migmatite derived from Proterozoic volcanic and sedimentary rocks \cite{Pi:09}. The total field anomaly shows magnetic targets elongated in the NE-SW direction. A data-space inversion algorithm with a Cauchy norm sparsity constraint on model parameters was applied by Pilkington \shortcite{Pi:09} for the inversion of this data set. Furthermore, the results of the inversion algorithm of  Li $\&$ Oldenburg \shortcite{LiOl:98} are also presented in Pilkington \shortcite{Pi:09}. Therefore, the results presented here can be compared with the results of presented in Pilkington \shortcite{Pi:09} and for  consistency we thus use a grid of $64 \times 64$ data points with $100$~m spacing and a uniform subsurface discretization of $64 \times 64 \times 20 = 81920$ blocks. The intensity of the geomagnetic fields, the inclination and the declination are $60000$~nT, $78.5^{\circ}$, $5.3^{\circ}$, respectively. As for the simulations we  set $\Kmax=50$, $\bfma=\mathbf{0}$, and impose bound constraints on the model parameters. In this case, these are $0=\kappa_{\mathrm{min}} \le \kappa \le \kappa_{\mathrm{max}}=0.2$ SI unit \cite{Pi:09}. The values of parameter $q$ are selected based on the presented rules in Section~\ref{synthetic}. We select  $q=1100>m/4$ and $q=2100>m/2$ with and without power iterations, respectively.

\begin{figure*} 
\subfigure{\label{fig20a}\includegraphics[width=.32\textwidth]{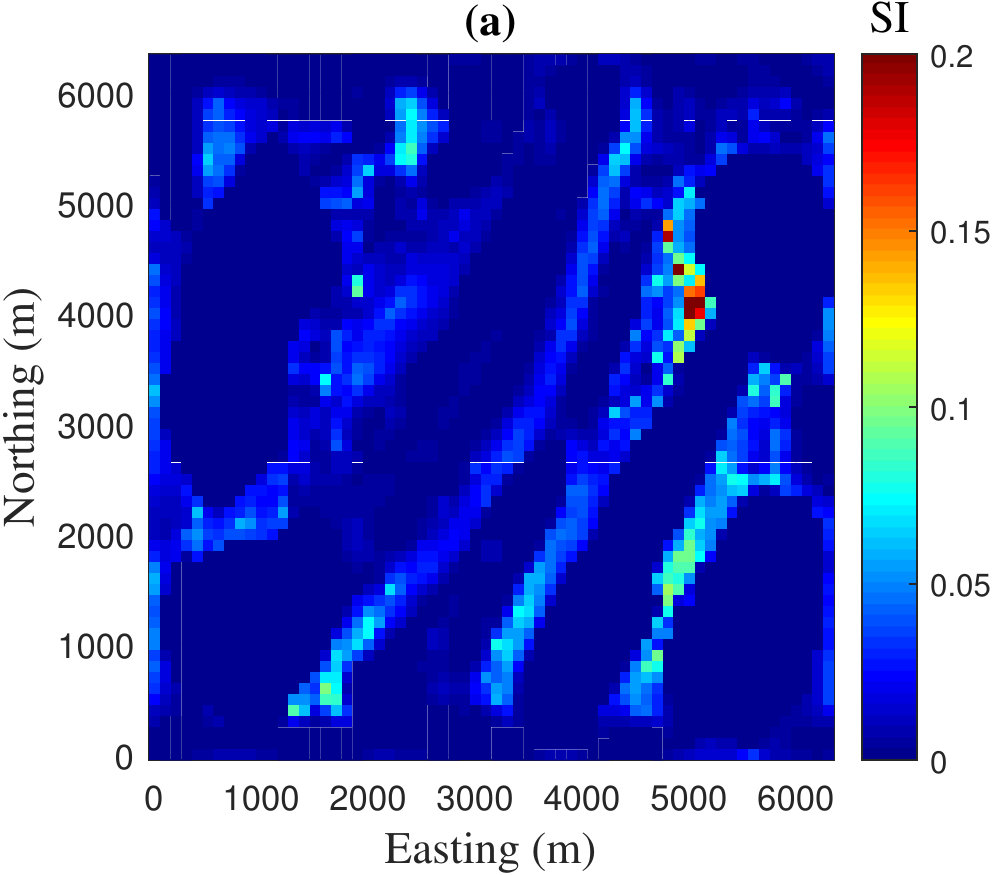}}
\subfigure{\label{fig20b}\includegraphics[width=.32\textwidth]{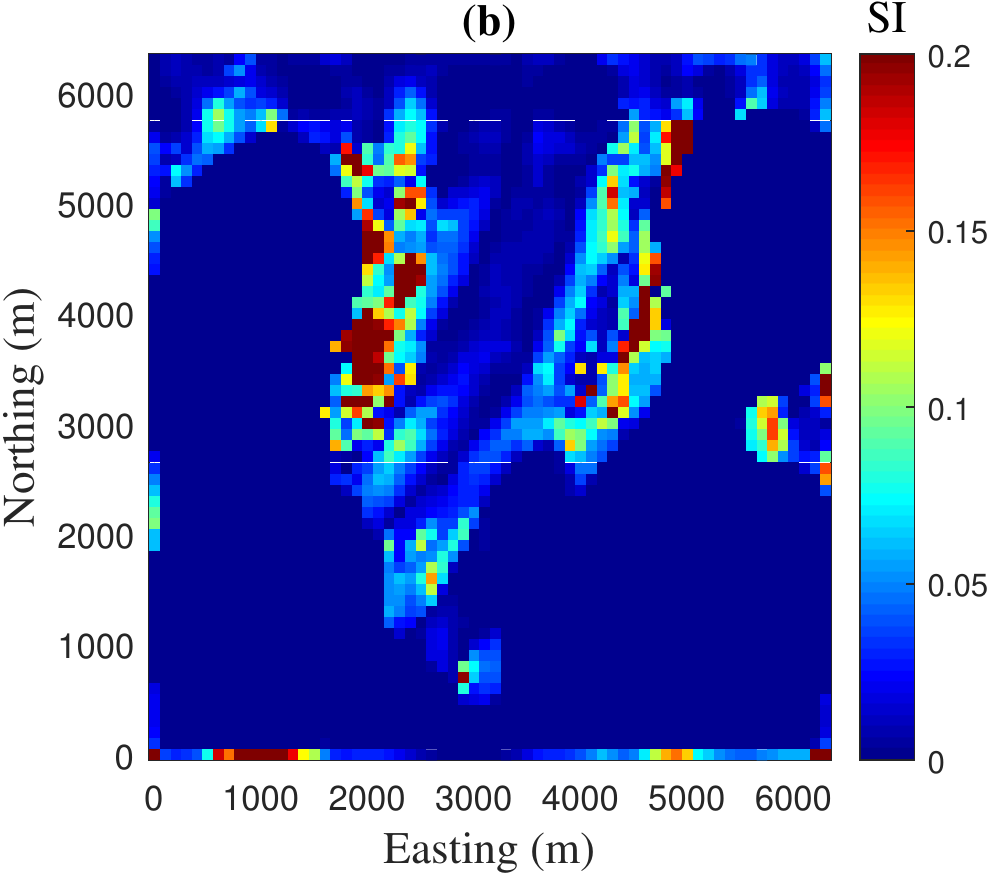}}
\subfigure{\label{fig20c}\includegraphics[width=.32\textwidth]{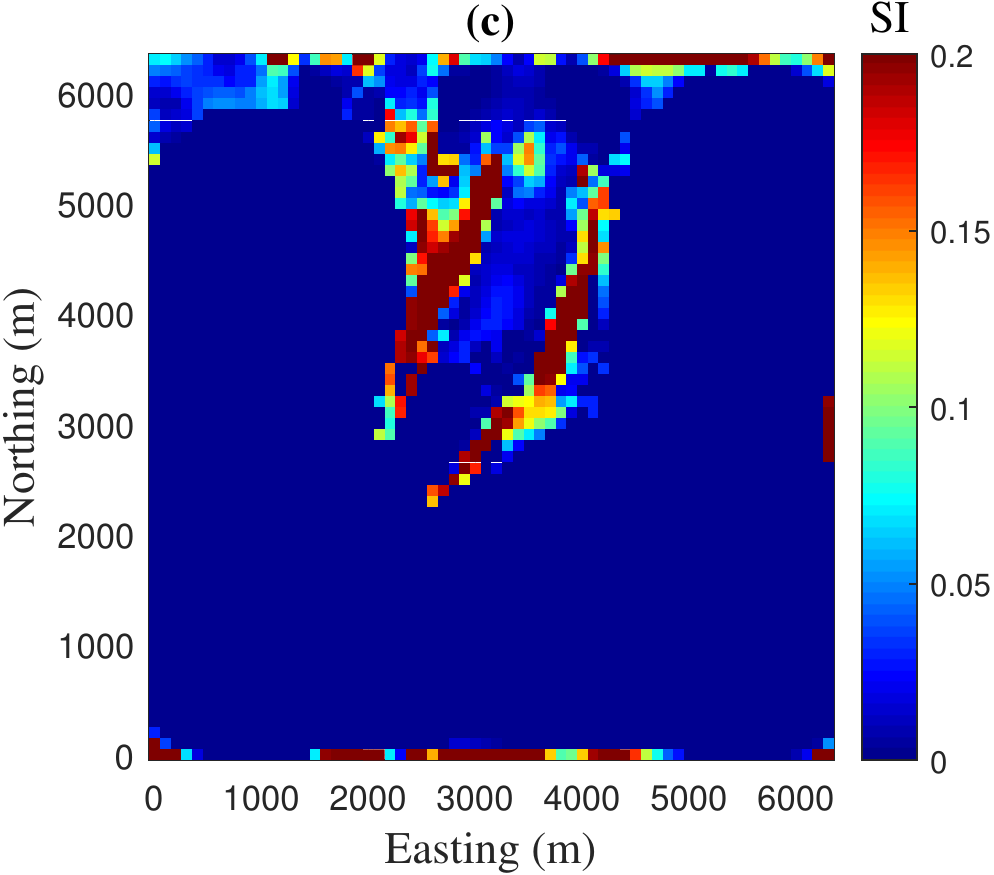}}
\caption {Reconstructed susceptibility model using \texttt{Hybrid-RSVD} via power iterations with $s=1$ and target rank $q=1100$ for the inversion of magnetic data given in Fig.~\ref{fig19a}. The plane-sections illustrate depths: (a) $300-400$~m;  (b) $500-600$~m;  (c) $700-800$~m. } \label{fig20}
\end{figure*}

\begin{figure*} 
\subfigure{\label{fig21a}\includegraphics[width=.45\textwidth]{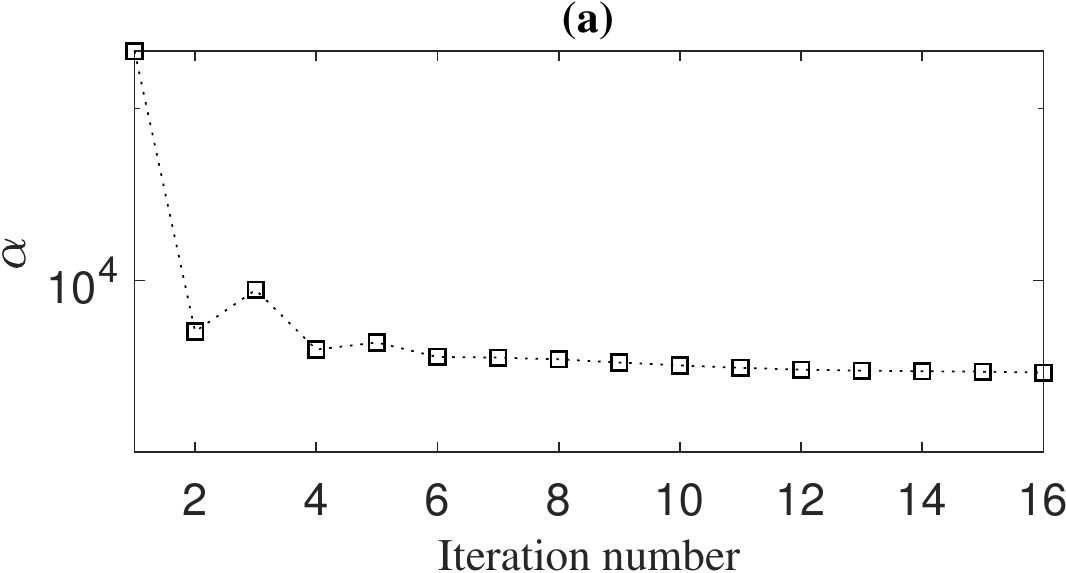}}
\subfigure{\label{fig21b}\includegraphics[width=.45\textwidth]{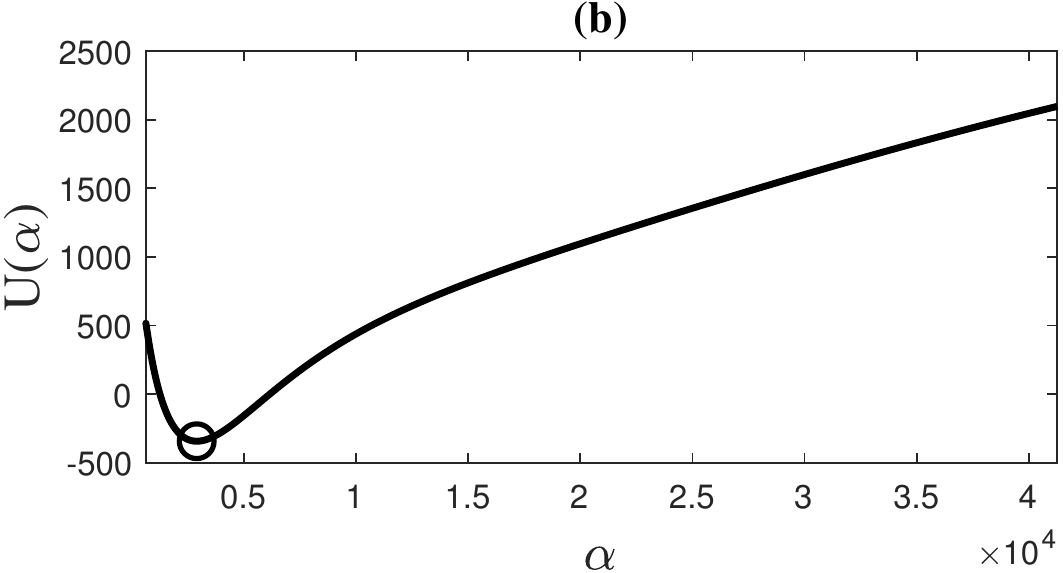}}
\caption { (a) The progression of regularization parameter $\alpha^{(k)}$ with iteration $k$; (b) The UPRE function at the final iteration. } \label{fig21}
\end{figure*}

\begin{figure*} 
\includegraphics[width=.8\textwidth]{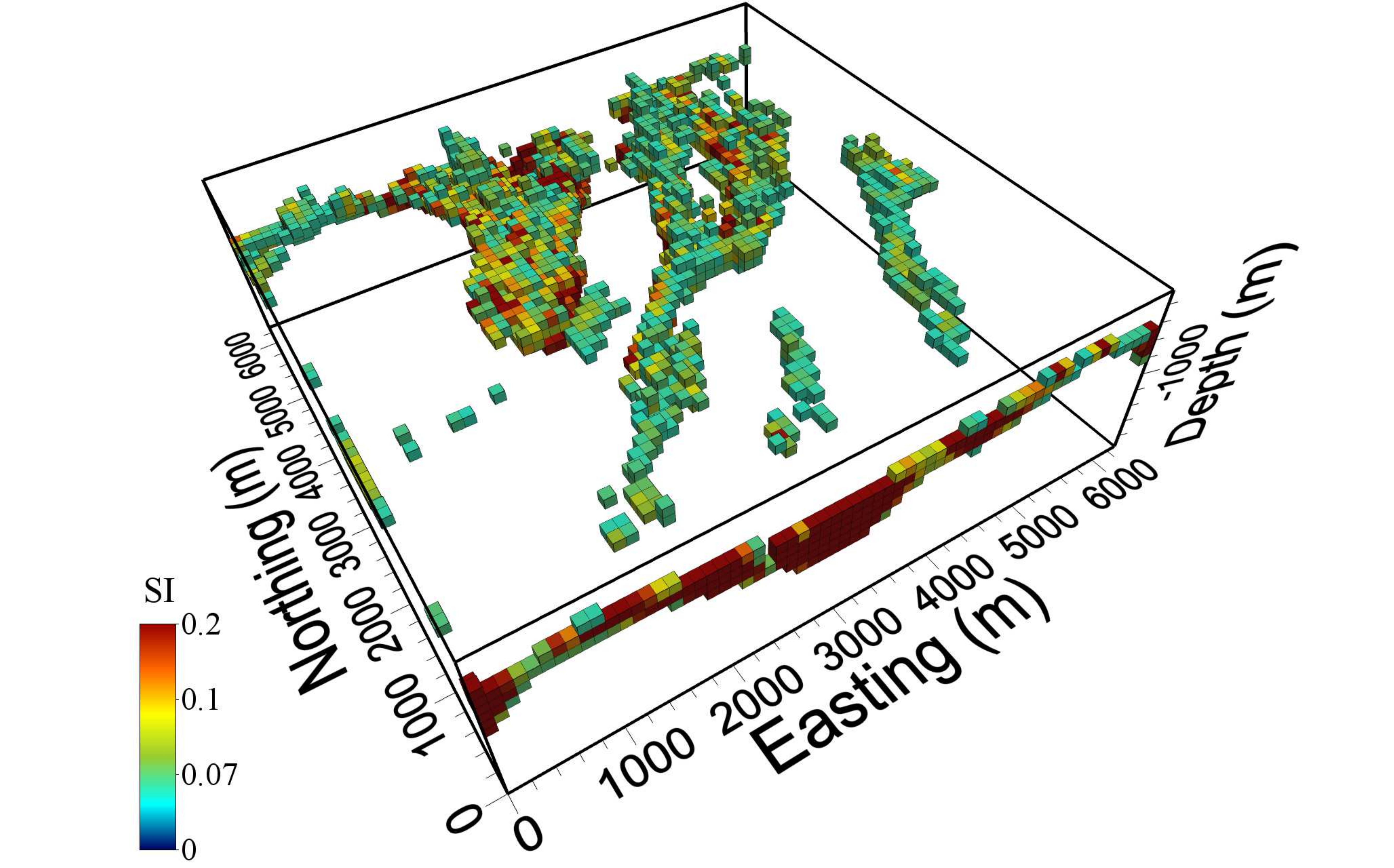}
\caption {$3$-D view of the reconstructed susceptibility model shown in Fig.~\ref{fig20}, illustrating cells with  $\kappa>0.05$ (SI unit).} \label{fig22}
\end{figure*}

The results of the inversions, as given in Table~\ref{realdatatab}, demonstrate that the methodology generates converged solutions using a limited number of iterations and at computational cost on the order of a few minutes only. Overall these results demonstrate the  feasibility of using the \texttt{Hybrid-RSVD} algorithm for the inversion of large-scale geophysical data sets. Three plane-sections of the reconstructed model obtained using $s=1$  for the power iterations are illustrated in Fig.~\ref{fig20}. The anomaly produced by this model is  shown in Fig.~\ref{fig19b}. The progression of the regularization parameter at each iteration and the UPRE functional at the final iteration are also presented in Fig.~\ref{fig21}. Furthermore,  Fig.~\ref{fig22} illustrates a $3$-D view of the model for cells with $\kappa>0.05$. Our results indicate that, generally, there are three main subsurface targets. The target in the South-East of the area starts from $300$~m and extends to $400$~m; it is not as deep as the other two targets. The target in the central part of the area, is elongated in the SW-NE direction, and starts from about $300$~m and extends to about $800$~m in depth. This target in its northeastern part is divided into two sub-parallel targets. The third main target, located in the central north part of the area, is the deepest target which starts at about $400$~m and extends to about $900$~m. The results in the shallow and intermediate layers are in agreement with the results of Pilkington \shortcite{Pi:09}, but at depth the  results presented here are more focused.

\begin{table}
\caption{Results of the inversion algorithms applied on the  magnetic data of Fig.~\ref{fig19a}.}\label{realdatatab}
\begin{tabular}{c  c  c  c  c c c c c}
\hline
 Method & $q$ &$\alpha^{(1)}$& $\alpha^{(K)}$& $K$ & $\chi^2$  & Time (s) \\ \hline
\texttt{Hybrid-RSVD} & $2100$ & $262575$ & $3630.2$ & $15$ & $4007.1$ & $559.7$ \\ \hline
\texttt{Hybrid-RSVD} with power iterations & $1100$ & $215644$ & $2906.7$ & $16$ & $4089.7$ & $464.0$ \\
\hline  
\end{tabular}
\end{table}

\section{Conclusions}\label{conclusion}
We present an algorithm for fast implementation of  large-scale focusing inversion of both magnetic and gravity data. The algorithm is based on a combination of the $L_1$-norm regularization strategy with the RSVD. For large-scale problems, the powerful concept of the RSVD provides an attractive and indeed fast alternative to methods such as the LSQR algorithm.  Here we have presented a comprehensive comparison of the \texttt{Hybrid-RSVD} methodology with power iterations for the inversion of  gravity and magnetic data. Generally, we have shown that there is an important difference between gravity and magnetic inverse problems when approximating a $q$-rank matrix from the original matrix. For the inversion of magnetic data it is necessary to take  larger values of $q$,  as compared with the inversion of gravity data, in order that a suitable approximation of the system matrix is obtained.  Furthermore, including power iterations within the algorithm improves the approximation quality. Indeed the RSVD obtained using the power iterations step yields approximation of the dominant singular space even for small choices of $q$. Thus, the RSVD with power iterations yields an efficient strategy when the singular values of input matrix decay slowly. In particular, the presented methodology  can be used for other geophysical data sets and that the choice of the rank $q$ approximation will depend on the spectral properties of the relevant kernel matrices. If the RSVD without power iteration does not approximate the dominant singular  values then power iterations should be included to improve the quality of estimated singular  values. Our results also demonstrate that it is possible to obtain a reasonable lower estimate for $q$, for both gravity and magnetic data inversions, based on the number of data measurements, $m$.  In conclusion, we have demonstrated that  it is feasible to use an efficient RSVD methodology for problems that are too large to be handled using the full SVD.  Application of the RSVD for the  joint inversion of gravity and magnetic data  is a topic for future work.

\begin{acknowledgments}
The authors would like to thank Dr. Mark Pilkington for providing data from Wuskwatim Lake area.  This study is supported by the National Key R\&D Program (NO. 2016YFC0600109) and NSF of China (NO. 41604087 and 41874122).

\end{acknowledgments}

\appendix

\end{document}